\documentclass[10pt,a4paper]{article}
\usepackage[utf8]{inputenc}

\usepackage{graphicx}
\graphicspath{ {figs/} }

\usepackage{algorithm}
\usepackage{algpseudocode}

\usepackage{amsmath}

\DeclareMathOperator*{\argmin}{arg\,min}

\usepackage[table,xcdraw]{xcolor}

\usepackage{arxiv} 
\usepackage{hyperref}       
\usepackage{lineno} 
\usepackage[super]{natbib}
\usepackage{mciteplus}      
\usepackage{booktabs}       
\usepackage{amsfonts}       
\usepackage{nicefrac}       
\usepackage{enumitem}       
\usepackage{xcolor}
\usepackage{caption}        
\usepackage{subcaption}     

\title{Development and Evaluation of Conformal Prediction Methods for QSAR}
\author{
Yuting Xu$^{a}$, Andy Liaw$^{a}$, Robert P. Sheridan$^{b}$, Vladimir Svetnik$^{a}$\\
\footnotesize $^{a}$ Biometrics Research, Merck \& Co., Inc., Rahway, New Jersey 07065, United States\\
\footnotesize $^{b}$ Computational and Structural Chemistry, Merck \& Co., Inc., Kenilworth, New Jersey 07033, United States \\
}
\date{\today}

\begin{document}

\maketitle


\begin{abstract}
The quantitative structure-activity relationship (QSAR) regression model is a commonly used technique for predicting biological activities of compounds using their molecular descriptors. Predictions from QSAR models can help, for example, to optimize molecular structure;  prioritize compounds for further experimental testing;  and estimate their toxicity.  In addition to the accurate estimation of the activity, it is highly desirable to obtain some estimate of the uncertainty associated with the prediction, e.g., calculate a prediction interval (PI) containing the true molecular activity with a pre-specified probability, say 70\%, 90\% or 95\%. The challenge is that most machine learning (ML) algorithms that achieve superior predictive performance require some add-on methods for estimating uncertainty of their prediction. The development of these algorithms is an active area of research by statistical and ML communities but their implementation for QSAR modeling remains limited.  Conformal prediction (CP) is a promising approach. It is agnostic to the prediction algorithm and can produce valid prediction intervals under some weak assumptions on the data distribution. We proposed computationally efficient CP algorithms tailored to the most advanced ML models, including Deep Neural Networks and Gradient Boosting Machines. The validity and efficiency of proposed conformal predictors are demonstrated on a diverse collection of QSAR datasets as well as simulation studies.
\end{abstract}


\section{Introduction}

Quantitative structure–activity relationship (QSAR) regression models are routinely applied computational tools in the drug discovery process for predicting the biological activities of molecules from the molecular structure-based features. 
The predictions are usually used to prioritize candidate molecules for future experiments and help chemists gain better understanding of how structural changes affect activities \cite{cherkasov2014qsar,muratov2020qsar,xu2022deep}. 
While most of the previous efforts focus on improving the accuracy of point predictions, quantifying the uncertainty in the prediction will add valuable insight\cite{sahlin2013uncertainty,sheridan2015relative,hirschfeld2020uncertainty,mervin2021uncertainty,yu2022uncertainty}.   
For regression tasks, prediction intervals are often used as quantitative measures of the reliability or confidence in the point prediction at a given probability. 
A well-calibrated prediction interval contains future observations with a prespecified probability, which is also called nominal coverage. The width of a calibrated prediction interval gives users an intuitive estimate of the prediction error magnitude, since a prespecified percent of the prediction errors is contained within the interval range. 
For example, an interval of 2.0 to 3.0 at 95\% probability for compound i means the true value of the activity has a 95\% chance of falling within 2 and 3.  In comparison, a wider interval of 1.0 to 4.0 at 95\% for compound j  would indicate that j is less reliably predicted than i.

There are a variety methods of estimating prediction uncertainties for QSAR predictions. 
Some methods directly estimate the quantile or variance of prediction errors, while others provide relative uncertainty scores between different molecules which require further calibration for obtaining valid prediction intervals\cite{hirschfeld2020uncertainty}.
Quantile regression methods\cite{koenker2001quantile,hao2007quantile} have been explored with various QSAR algorithms, including Neural Networks\cite{el2014quantile}, Random Forests\cite{feng2019building}, and Light Gradient Boosting Machine\cite{difranzo2020nearest}; however, the prediction intervals constructed from the quantile estimates often lead to an under-coverage, i.e., the actual coverage of these intervals is smaller than that of the nominal coverage specified by a user. 
The Bayesian methods estimate the posterior distribution of molecular activity given a molecular structure. Thus, the prediction intervals could be computed using the variance of the prediction errors under certain assumption of their distribution. An example of this approach is studied in Dai et al. \cite{feng2019building}, the Bayesian Additive Regression Trees that simultaneously estimate molecular activity and the error variance as a measure of prediction uncertainty, under the assumption that the errors are normally distributed. This assumption, however, is often violated in many QSAR applications. 
A practical concern with some methods is the high computational cost, which is especially high for Bayesian methods. 
Hirschfeld et al. \cite{hirschfeld2020uncertainty} benchmarked several uncertainty quantification methods that are applicable to Neural Networks, including ensemble-based methods\cite{cortes2018deep,cortes2019reliable}, distance-based methods, mean-variance estimation\cite{nix1994estimating}, and union-based methods\cite{huang2015scalable}. This work recommended several top-performing methods that produced an estimate of the predicted error variances with the lowest negative log likelihood under the Gaussian distribution assumption for prediction errors. The likelihood-based evaluation metric, however, is highly sensitive to the distributional assumption. As was mentioned above, prediction errors in many QSAR applications are non-Gaussian. The top recommended methods in Hirschfeld et al. \cite{hirschfeld2020uncertainty} are union-based methods involving training a neural network for point prediction, and a second model, for example, the Gaussian Process or Random Forest regression to estimate the variance of errors in prediction produced by the first model. Training of two models typically requires substantially more data. 
Comprehensive reviews on uncertainty quantification methods and/or applications could be found in Mervin et al.\cite{mervin2021uncertainty}, Abdar et al. \cite{abdar2021review}, Tian et al. \cite{tian2022methods} and Yu et al.\cite{yu2022uncertainty}.

We are seeking practical methods to provide prediction intervals accompanying the point predictions for large QSAR datasets in an industrial-scale pharmaceutical drug discovery environment. 
A good algorithm should meet the following requirements:
\begin{enumerate}[noitemsep]
    \item \emph{Marginal validity.} The marginal coverage probability of the prediction interval should not be less than a user-specified (nominal) coverage level. Marginal, is roughly understood as averaged over all molecules. The empirical marginal coverage for a test set is the proportion of prediction intervals that contains these molecules' true activity measurements. 
    \item \emph{Conditional validity (informativeness, or adaptivity).} The conditional coverage, i.e. prediction interval coverage for each molecule,  should be close to the nominal coverage. Roughly speaking, the width of the prediction interval should closely follow the uncertainty in the prediction of any molecule. For example, with a homoscedasticity where the variance of prediction errors does not depend on the molecular descriptors, prediction intervals for all molecules will have the same width. In contrast, with heteroscedasticity, where the variance depends on the descriptors, the width of the prediction interval should increase or decrease following the variance. 
    Compared to marginal validity, conditional validity is a much more stringent requirement usually impossible to achieve. However, we still hope to have prediction intervals with sizes adaptive to the prediction errors, in order to approximate this property. 
    \item \emph{Efficiency, or tightness of prediction intervals.} The width (length, or size) of the prediction intervals should be as small as possible. 
    \item \emph{Low computational cost.} Preferably, the computation of prediction intervals should not involve a significant computational effort in addition to training the point prediction model. 
    \item \emph{Independence of model or data distribution.} The properties of the prediction intervals, such as validity, efficiency, and adaptivity, do not depend on the choice of point prediction model and the probability distribution of data or prediction errors.
\end{enumerate}

Many evaluation metrics of prediction intervals have been explored in previous benchmark studies, and several recommendations have been proposed \cite{khosravi2011comprehensive,scalia2020evaluating,hirschfeld2020uncertainty,tian2022methods}. 
Although it is difficult to find a single method that is superior to others under all criteria, the conformal prediction (CP) framework could satisfy most of these practical considerations and has attracted increased attention in recent QSAR applications\cite{eklund2012application,carlsson2014aggregated,norinder2014introducing,eklund2015application,svensson2018conformal,cortes2018deep,cortes2019concepts,cortes2019reliable}. 
The prediction intervals generated by conformal prediction methods are guaranteed to achieve valid marginal coverage, i.e. on average the test set prediction intervals to contain the measured molecular activities with a user-specified probability, under the assumption of data exchangeability\cite{vovk2005algorithmic,lei2018distribution}. This marginal validity property does not depend on any distributional assumptions or model assumptions, and holds for any sample data size\cite{angelopoulos2021gentle}. 
The conformal prediction is applied as a companion to any pre-trained model for a QSAR regression task, and many conformal algorithms require little extra computational costs besides the training of the underlying point predictor. 
Construction of conformal prediction intervals is based on a ``nonconformity'' score that quantifies how unusual a data point is relative to the training data\cite{shafer2008tutorial}. We prefer to use nonconformity scores that do not require additional modeling efforts. The adaptivity property of prediction intervals depends on the choice of nonconformity score.
A theory for selecting the nonconformity score is yet to be developed, and the choice needs to be evaluated with empirical studies. 
Several computationally efficient conformal prediction algorithms have been developed for QSAR regression\cite{svensson2018conformal,cortes2018deep,cortes2019reliable}, however, a comprehensive evaluation of their properties, especially their adaptivity, i.e., ability to handle heteroscedasticity of prediction errors, remains to be done. 

In this work, we develop a computationally efficient conformal prediction algorithm, named adaptive calibrated ensemble (ACE) that is based on a specifically designed nonconformity score. 
The ACE algorithm can be used with any model that provides both a point prediction and an uncertainty estimate for each molecule. 
For any ensemble-based model, the mean prediction over the ensemble is the ``prediction'' (i.e. the point estimate) and the standard deviation of predictions can be used as the prediction uncertainty estimate. The point estimate and prediction uncertainty estimate are inputs for ACE. 
The ACE nonconformity score is computed from the point prediction and the data-driven transformation of the prediction uncertainty score. 
If the uncertainty scores effectively represent the relative uncertainty between molecules, the ACE prediction intervals would achieve not only a pre-specified marginal coverage but also an approximate conditional coverage. 

Deep neural nets (DNN) and Gradient Boosting models (GB) are among the most predictive descriptor-based ML methods \cite{ma2015deep,svetnik2005boosting,sheridan2016extreme}. Our second contribution is to propose novel approaches for obtaining uncertainty estimates for these methods, called DNN-multitask and GB-tail, respectively. DNN-multitask is compared to the state-of-the-art DNN-dropout method \cite{gal2016dropout,yu2022uncertainty}.  

Our third contribution is a comprehensive analysis of a diverse collection of real QSAR datasets, and simulations derived from it, 
with special attention to the analysis of conditional validity and efficiency. 

The paper is organized as follows. 
Section \ref{sec: Datasets} describes QSAR datasets and their molecular descriptors. 
In Section \ref{sec: Methods} we briefly review the conformal prediction framework and applications to QSAR; introduce the proposed ACE conformal prediction algorithm; define evaluation metrics and explain in detail how to get ensemble predictions or raw prediction uncertainty scores for several state-of-the-art machine learning algorithms. 
Section \ref{sec: Simulation} contains benchmarking results obtained on simulated data with either homoscedastic or heteroscedastic errors. 
Section \ref{sec:Application} presents the results of applying ACE jointly with several popular QSAR predictive algorithms, and shows the validity, efficiency and adaptivity of proposed methods on diverse QSAR datasets.

\section{Data Sets and Molecular Descriptors}
\label{sec: Datasets}

Two sets of QSAR data collections are used in this study: 
\begin{itemize}
    \item ChEMBL: $IC_{50}$ data sets for 23 diverse protein targets and receptors from the ChEMBL database. The data sets were obtained from Cortes-Ciriano et. al. \cite{cortes2018deep}, excluding the smallest dataset ``A2a'' with only 203 molecules. We generated molecular descriptors according to the procedures described in Cortes-Ciriano et al. \cite{cortes2019reliable}: Compute the circular Morgan fingerprints \cite{rogers2010extended} using RDkit \cite{landrum2006rdkit} (version 2021.03.2) with the radius set to 2, and the output fingerprint length is 2048.
    \item Kaggle: The 15 QSAR datasets used in the 2012 ``Merck Molecular Activity Challenge'' Kaggle competition and released in Ma et al. \cite{ma2015deep}. These data sets are of various sizes for either on-target potency or off-target absorption, distribution, metabolism, and excretion (ADME) activities. The molecular descriptor is the union of the ``atom pair'' (AP) descriptors from Carhart et al. \cite{carhart1985atom} and ``donor–acceptor pair'' (DP) descriptors \cite{kearsley1996chemical}. Each data set was provided in two parts: the time-split training set and test set. In this work, we only use the training set in each data set, since investigating the covariate-shift problem for time-split test set is out of scope.    
\end{itemize}

\begin{figure}[htbp]
\caption{Dimension of two groups of data sets}
\centering
\includegraphics[width=10cm, height=6cm]{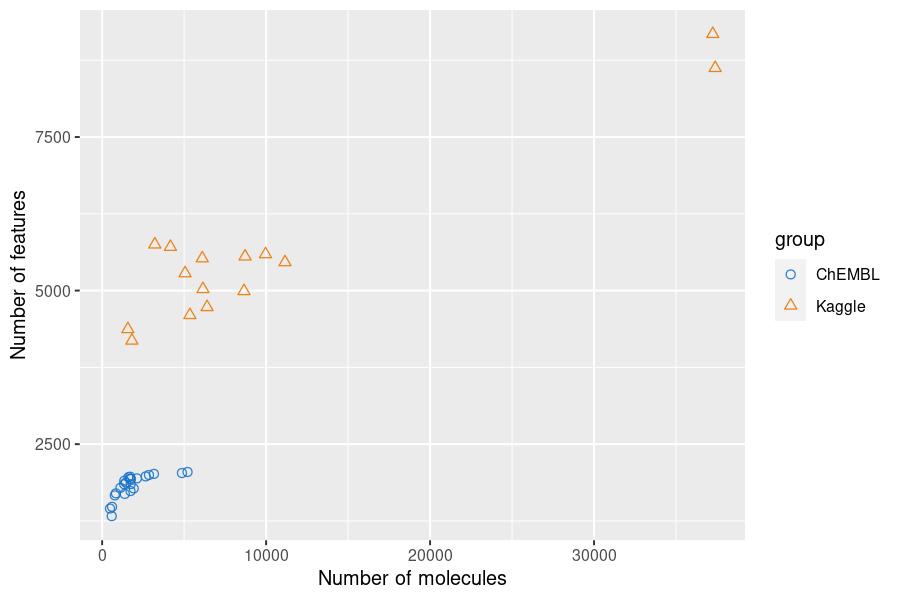}
\label{fig:datasets}
\end{figure}

\begin{figure}[htbp]
\centering
\caption{Distribution of activities for ChEMBL data sets. The molecular activities are log-scaled $IC_{50}$ values: $pIC_{50} = -log_{10} IC_{50}$}
\includegraphics[width=\textwidth]{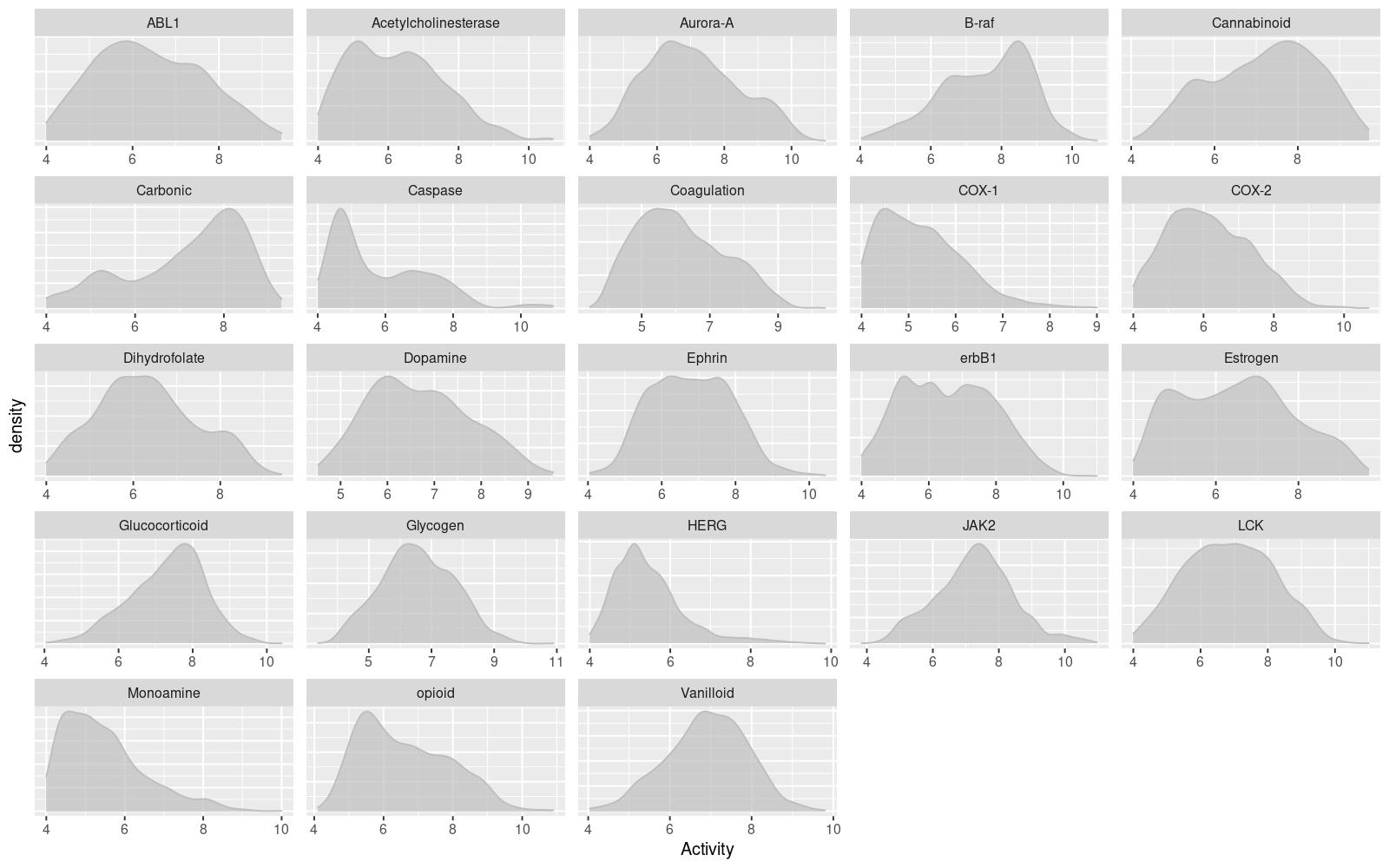}
\label{fig:datasets_act_ChEMBL}
\caption{Distribution of activities for Kaggle data sets. The molecular activities are scaled to zero mean and unit variance within each data set for better visualization.}
\includegraphics[width=\textwidth]{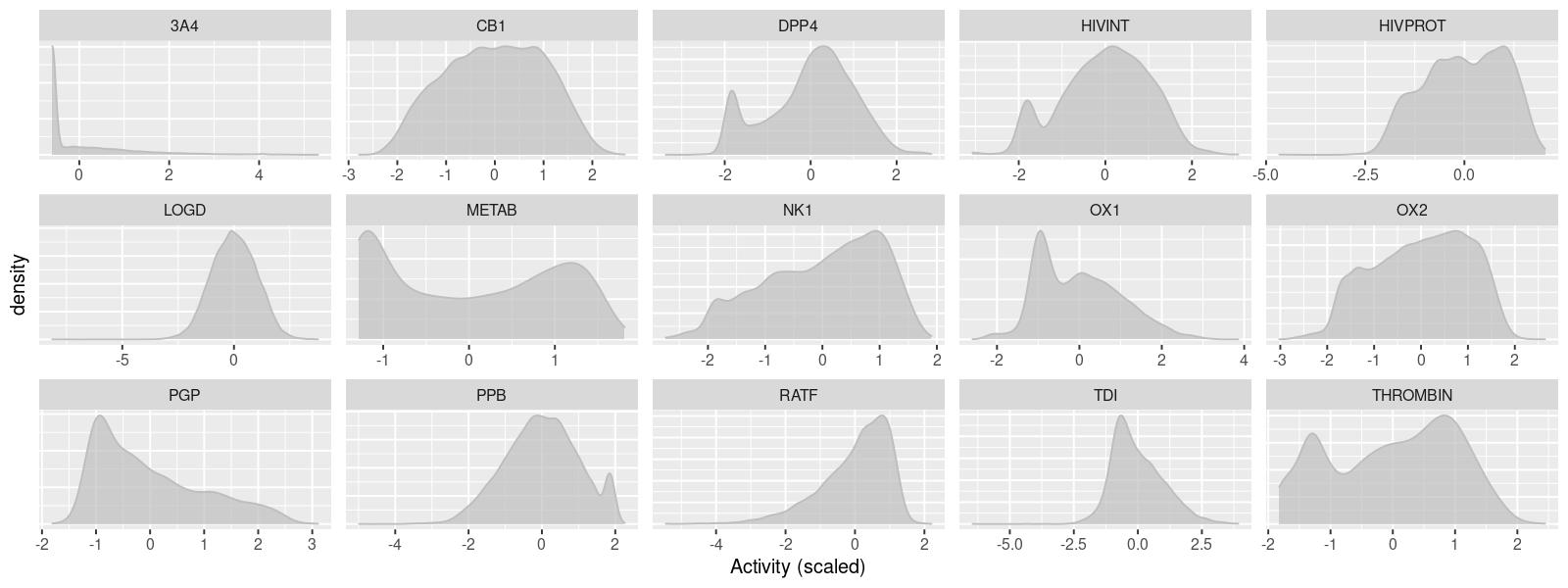}
\label{fig:datasets_act_Kaggle}
\end{figure}

Compared to the ChEMBL collection, the Kaggle collection contains a mixture of easy and hard to predict tasks, larger data sets, and more challenging distribution of activities. 
Figure \ref{fig:datasets} shows the distribution of the number of unique features versus number of molecules in these two QSAR data collections. 
Figure \ref{fig:datasets_act_ChEMBL} and Figure \ref{fig:datasets_act_Kaggle} contain the distribution of molecular activities in each dataset. 

Due to the complex nature of QSAR datasets, the distribution of prediction errors is often unknown. 
Some data sets may have near constant prediction error variance that is unrelated to molecular features. In this case, applying a constant width prediction interval for all molecules (i.e. homoscedastic) is preferred. Other datasets may have varying prediction error variance for different molecules, for example, higher randomness in experimental measurements for more active compounds. 
In this case, prediction intervals with varying widths correlated with prediction errors (i.e. heteroscedastic) are desirable. A good algorithm should provide adaptive prediction intervals suitable for both cases. However, the two groups of QSAR datasets are likely to contain a mixture of both cases. 
We also use simulated data with Gaussian distributed noises that have either constant variance or changing variance to demonstrate how  the methods performed in these cases.

\section{Methods}
\label{sec: Methods}

\subsection{Conformal Prediction}

Conformal prediction (CP) is a model-agnostic framework for measuring prediction uncertainty \cite{shafer2008tutorial,vovk2005algorithmic}.
In regression tasks, the prediction intervals constructed by conformal prediction achieve valid marginal distribution-free coverage with only the general i.i.d. (independent and identically distributed) data assumption \cite{lei2018distribution,papadopoulos2008inductive}.
The original CP algorithm, which is called transductive conformal prediction, is computationally expensive since it requires training a new predictive model for every test set sample. For large datasets in QSAR problems, a modified version of the CP algorithm called Inductive Conformal Prediction (ICP) or Split Conformal Prediction have been used \cite{papadopoulos2002inductive,eklund2015application,svensson2018conformal,cortes2018deep,cortes2019reliable}. The steps are:

\begin{enumerate}
    \item Specify a machine learning method to generate a model from which we get point predictions.
    \item Define a nonconformity measure to quantify how unusual a sample is compared to the rest of the samples. The nonconformity measure may depend on the prediction $\hat{y}$ or the feature $\mathbf{X}$.  
    \begin{itemize}
        \item For constructing homoscedastic prediction intervals, which have a constant width for all molecules, the nonconformity measure is defined as the absolute value of residuals: 
            \[ \alpha = |y - \hat{y}| \]
        \item For constructing heteroscedastic prediction intervals that apply to individual molecules, the nonconformity measure can be defined as the normalized absolute value of residuals:
            \begin{equation}
            \label{eq:1}
                \alpha = \frac{|y - \hat{y}|}{\sigma(\mathbf{X})}
            \end{equation}
        where $\sigma(\mathbf{X})$ is an estimate of the accuracy of the point predictor\cite{johansson2014regression}. The homoscedastic nonconformity measure is corresponding to a special case when $\sigma(\mathbf{X})\equiv 1$. 
    \end{itemize}
    \item Divide the available training data randomly into a ``proper training set'' and a ``calibration set''. For QSAR tasks, the relationship between molecular structure descriptors and activities is usually highly complex. It is preferable to use large fraction of data, e.g. around 80\%, as the ``proper training set''.   
    \item Use the proper training set to construct a model, and use the model to predict the activity of molecules in the calibration set. 
    Calculate the nonconformity score $\alpha_j$ for each calibration set sample $x_j~(1\leq j \leq n)$. 
    \item Specify a desired nominal coverage probability $\theta$, and calculate the $\theta^{th}$ percentile of nonconformity scores in calibration set as $\alpha_{\theta}$. 
    \item The prediction interval for a new molecule $\mathbf{x}_{new}$ in the test set would be: 
        \[\hat{y}(\mathbf{x}_{new}) \pm \alpha_{\theta} \sigma(\mathbf{x}_{new})\]
\end{enumerate}

The choice of the scaling factor $\sigma(\mathbf{X})$ in nonconformity measure affects the efficiency of conformal prediction (i.e. the width of prediction intervals) \cite{shafer2008tutorial}. There are two typical approaches how to define the scaling factor used in previous studies: model-based \cite{papadopoulos2002inductive,johansson2014regression,eklund2015application,romano2019conformalized} and ensemble-based \cite{svensson2018conformal,cortes2018deep,cortes2019reliable}. 
In this paper we will use the ensemble-based approach. Here $\sigma(\mathbf{X})$ is a function of the ensemble prediction uncertainty $s(\mathbf{X})$, which is the standard deviation of predictions from an ensemble model \footnote{Below we abbreviate $\sigma(\mathbf{X})$ and $s(\mathbf{X})$ as $\sigma$ and $s$ respectively, when they are clearly depended on molecular feature $\mathbf{X}$.}. 
Ensemble-based scaling factors performed slightly better than the model-based scaling factors in previous work on QSAR datasets \cite{svensson2018conformal}.  Some machine learning methods like random forest natively produce an ensemble of predictions. Other popular machine learning methods such as DNN or Boosting need modification to produce an ensemble of predictions, hopefully without being too computationally expensive. This is discussed in Section \ref{sec: Methods-ML}.

\subsection{Adaptive Calibrated Ensemble (ACE) Conformal Predictor}

We need to transform $s$, the ensemble prediction uncertainty, to be used as the scaling factor $\sigma$ in equation \ref{eq:1}.
The exponential transformation of $s$, i.e. $\sigma = e^{s}$, has been a popular choice of the scaling factor in previous studies \cite{papadopoulos2011regression,svensson2018conformal,cortes2018deep,cortes2019reliable}. We call this scaling method ``expSD''.
Svensson et al. \cite{svensson2018conformal} explored different transformations $ \sigma = e^{\gamma \times s}$ with various weight $\gamma$ ranging from 0.05 to 1.25, and concluded that the best average performance was obtained when $\gamma = 1$. 
However, in our experience, this scaling method is not always optimal for different datasets or ensemble approaches. The range of scaling factor $\sigma$ affects the range of PI widths in a dataset. For ensemble methods that produce smaller $s$, the exponential transformation will lead to near constant scaling factor $\sigma$ and thus almost constant-width prediction intervals, which may not be desirable if the dataset is intrinsically heteroscedastic. It may also lead to excessively wide and non-informative prediction intervals for some molecules.   

We proposed a flexible calibration algorithm, Adaptive Calibrated Ensemble (ACE), which will generate transformation of any raw prediction uncertainty score $s$ as a scaling factor for nonconformity scores adaptive to different datasets and/or ensemble models. 
Here $s$ could be some measure of the variability of predictions in an ensemble of models, or any relative uncertainty scores that correlate with the prediction errors. 
The ACE algorithm finds the optimal transformation by minimizing the  coverage error conditional on the PI width via repeated cross-validation on the calibration set. The major steps in ACE are:
\begin{enumerate}
    \item Calculate the mean $\mu_s$ and standard deviation $sd_s$ of the $s$ values on the calibration set, and calculate the normalized $s$ as $\tilde{s} = \frac{s - \mu_s}{sd_s}$.
    \item Define $b$ as the average of the absolute error in the validation set. And define the scaling factor $\sigma$ as a function of the parameter $a$:
        \[ \sigma(a;\tilde{s}):= a*\tanh(\tilde{s})+b \]
        where $0 \leq a \leq b$ to ensure $\sigma(a)$ is always a positive and non-decreasing function of $\tilde{s}$.
    \item Perform repeated two-fold cross-validation on the calibration set, and use grid-search to find the optimum value of $a$ to achieve the lowest average coverage error over four equal size subgroups defined by prediction interval widths.
\end{enumerate}

More details are provided in Algorithm \ref{alg:ACE} in Supporting Information Section \ref{SI:pseudocode}.

\subsection{Machine Learning Algorithms}
\label{sec: Methods-ML}

In this section, we describe the supervised learning algorithms commonly used in QSAR applications, and how to generate computationally efficient prediction uncertainty score $s$.
Additional details of implementation and hyperparameter settings of each algorithm were provided in the Supporting Information Section \ref{SI:Implementation}. 

\subsubsection{Random Forests}

Random Forest (RF) has been a very popular QSAR method due to its high prediction accuracy and robustness to the choice of hyper-parameters \cite{svetnik2003random}. RF is an ensemble of independent decision trees. The average of individual tree predictions is the final prediction for a molecule; and $s(X)$ is the standard deviation of tree predictions of the molecule represented by $X$, which has been an effective measure of prediction uncertainty in previous conformal applications \cite{svensson2018conformal,cortes2018deep,cortes2019reliable}. In these referenced reports, the expSD scaling method was used to define the nonconformity score and we use it here as well. 

Compared to other supervised learning methods, RF has a unique appealing feature: the out-of-bag (OOB) data already represents a random split, and one does not need to make an explicit ``proper training set/calibration set'' split\cite{johansson2014regression}.  
This prediction uncertainty estimation strategy is referred to as ``RF-OOB'' in the later sessions. 

\subsubsection{Deep Neural Networks}
Deep neural network (DNN) is a practical QSAR method for large datasets that usually achieve superior predictive performance \cite{ma2015deep}, although it is somewhat expensive computationally and sometimes requires GPU computing hardware. Here we consider only fully-connected descriptor-based DNNs.

We considered two approaches for generating ensemble predictions that require training only one DNN prediction model: the ``DNN-dropout'' as it is called in previous reports \cite{gal2016dropout,cortes2019reliable} and a novel ``DNN-multitask'' method.

\begin{itemize}
    \item DNN-dropout: Cortes Ciriano et al. \cite{cortes2019reliable} built conformal predictors for the ChEMBL datasets using test-time dropout \cite{gal2016dropout} ensembles, where the ensemble prediction was created by allowing random dropout of DNN hidden layers nodes and repeating the prediction process 100 times. The average and standard deviation across the 100 repeated predictions for a molecule were used as its predicted value and a measurement of the prediction uncertainty respectively. 
    
    We adopted the same DNN structure and optimization algorithm described in \cite{cortes2019reliable} for the ChEMBL data sets. 
    In Cortes Ciriano et al. \cite{cortes2019reliable}, multiple dropout rates, 0.1, 0.25, and 0.5, were explored, and the results demonstrated that the performance for different dropout rates was comparable. We adopted the median value of the dropout rate, i.e. using a constant dropout rate of 0.25 in all hidden layers.
    For Kaggle datasets, we used the DNN parameter setting from Ma et al. \cite{ma2015deep}.
    
    \item DNN-multitask: We create an artificial sparse multi-task DNN model from a single-task QSAR dataset, by replicating the outcome molecular activities K times with random omissions of molecules (with the probability of omission p), e.g.: 
        \[ \tilde{y}_i = (y_i, y_i, NA, y_i, NA, ...), \forall i \]
    where the artificial vector outcome $\tilde{y}_i$ for each molecule i is a K-dimensional vector with elements $\tilde{y}_{ij}, 1\leq j \leq K$,  
    \[Pr(\tilde{y}_{ij} = y_i) = 1-p, \qquad Pr(\tilde{y}_{ij} = NA) = p \]
    ``NA'' represents an omitted molecule. The average of multi-task outputs is taken as the final prediction. 
    The standard deviation of multi-task predictions for each molecule is used as the prediction uncertainty metric $s(X)$. The intuition is that molecules that are ``easier'' to predict would get more consistent predictions across this pseudo-ensemble model.  
    To explore whether the multi-task standard deviation is sensitive to the choice of hyperparameters K and p, we evaluated multiple combinations of $(K, p)$ pairs via simulations (see Supporting Information Section \ref{SI:tuning}), and recommend using a higher omissions probability (p=0.6), and a moderate number of the output nodes (K=20 or 50). 
    
    For Kaggle datasets, the neural network structure and optimization algorithm are the same as Ma et al. \cite{ma2015deep}, except for the output layer size. For the ChEMBL datasets, since the dimensions of input molecular features are smaller and the molecules are ``easier'' to predict, we use a smaller neural network structure, which achieved similar prediction accuracy with less computational cost compared to the DNN-dropout setting in Cortes Ciriano et al. \cite{cortes2019reliable}. 
\end{itemize}

\subsubsection{Gradient Boosting}

 Gradient  Boosting (GB) is a widely used QSAR method due to its computational efficiency and accuracy \cite{svetnik2005boosting,sheridan2016extreme}. A gradient boosting model consists of a series of decision trees.  The prediction of the model on a molecule is the sum of predictions of the decision trees on that molecule. The trees are added to the model in iterations such that the errors from the current model are used to grow a new tree, so that the new tree is forced to learn information that the current model did not learn from the data. Thus, in general we expect the trees learned earlier in the sequence to have larger contributions to the overall prediction than those later in the sequence.  For prediction of a new molecule, if the quality of the prediction is high, we would expect that contribution of trees in the sequence diminish for later iterations. If the magnitude of contribution to the prediction of a molecule falls to nearly  zero after the first few trees, we would say that the prediction for that molecule is reliable. On the other hand, if the magnitude of the contributions from the last few trees to the prediction is still large, then the prediction is less likely to be reliable. Therefore, we propose the ``GB-Tail'' method: for each molecule, let s be the mean absolute  value of the contributions from the final  w fraction of trees (for instance for $w=0.2$, the last 200 out of 1000 trees). This approach requires only post-processing of tree predictions and has a minimal computational cost.  

To check whether the estimate of the prediction uncertainty by the GB-tail method is sensitive to $w$, 
we investigated the impact of varying the hyperparameter $w$ (Supporting Information Section \ref{SI:tuning}), and recommend $w=0.2$.

\subsection{Evaluation Metrics}

We compared three types of conformal prediction intervals: Homoscedastic prediction intervals (homo), heteroscedastic prediction intervals with the ``expSD'' scaling factor (expSD), and heteroscedastic prediction intervals with scaling factor computed from the proposed ACE algorithm (ACE). They are applied in combination with four supervised learning algorithms/ensemble schemes introduced in Section \ref{sec: Methods-ML}.

We evaluated multiple important aspects of the prediction models, including the accuracy of point predictions, computational costs, and the informativeness of prediction intervals. All the evaluation metrics are calculated on the test set and average across multiple random repeats and/or datasets.  

The prediction accuracy is measured by the squared Pearson correlation coefficient (R-squared) and the normalized root mean squared error (RMSE). We would like to first ensure the prediction accuracy from different machine learning methods are satisfactory before evaluating the conformal prediction intervals. 

The performance of a conformal predictor is usually evaluated in terms of validity and efficiency. 
A valid prediction interval has a coverage probability no less than the prespecified nominal coverage level, i.e. if the probability is 90\%, no less than 90\% falls within the interval, and the efficiency is measured by the width of the prediction interval. In regression, a conformal predictor is preferable if it gives narrow intervals around the point predictions while achieving the target nominal coverage. 
The prediction intervals by any ICP algorithms are guaranteed to achieve marginal validity under any data distribution, but the choice of nonconformity score will affect the efficiency \cite{shafer2008tutorial,papadopoulos2008inductive}. 

For each nominal coverage probability, we calculate the coverage error as the difference between the test set coverage probability and the nominal coverage. For example, if the nominal coverage of the interval is 95\% and the actual coverage of the interval is 85\%, the difference is 10\%. 
Although ICP theoretically guarantees the coverage is correct, 
the actual finite sample test set coverages will in practice fluctuate around the nominal value \cite{angelopoulos2021gentle}.
We also calculated the mean absolute error of test set coverage across different datasets and repeats.  
To compare the efficiency across algorithms and datasets, one needs to normalize the raw prediction intervals. One way to do this is to find the ratio of the width of the raw interval to the width of the interval in the corresponding ``no-model'' case. The no-model case uses the distribution of observed training set activities rather than predictions. For example for a nominal coverage of 90\%, the interval is between the 5th-percentile and 95th-percentile values of the observed activities in training set. 
The ``no-model'' case interval should be wider than those obtained by the ICP analysis where the model-based predictions are used. 

When comparing the efficiency between heteroscedastic and homoscedastic prediction intervals, we scaled the heteroscedastic prediction interval widths by the corresponding homoscedastic prediction interval width for the same test set and the same point prediction model. 
The ability to differentiate easier from more difficult to predict molecules, i.e. ``adaptability'', is another top consideration for constructing informative prediction intervals \cite{angelopoulos2021gentle}. 
The expectation is that molecules which can be more accurately predicted have narrower prediction intervals while both prediction intervals achieve the nominal coverage probability. Ideally, one would like the prediction interval covers the true activity with the prespecified nominal coverage probability for every molecule. 
However, this cannot be achieved without imposing assumptions on the data distribution. 
Although there is no theoretical guaranteed for ICP algorithms to satisfy the conditional validity, one may approximate this property with heteroscedastic prediction intervals constructed from well-designed nonconformity scores. We used a size-stratified coverage metric to evaluate the adaptivity of prediction intervals in the following way. First, we sorted the test set molecules by their prediction interval widths, and divided them into five equal-size subgroups. We then calculated the coverage error within each subgroup, as well as the average absolute error of coverages across five subgroups for each dataset. If the coverage in these subgroups significantly deviates from the nominal coverage, it indicates a greater violation of the conditional coverage \cite{angelopoulos2021gentle}.

\section{Results}
\label{sec: Results}

In all the following numerical experiments, each data set is randomly split into a proper training set (70\% of the data) for training a prediction model for molecular activity and generating raw uncertainty scores; a calibration set (15\%) for conformal prediction; and a test set (15\%) for evaluation. The data split is repeated 20 times, and the same split is used for all algorithms. For RF models, since we are using the OOB data for conformal prediction instead of a separate calibration set, the union of the proper training set and calibration (85\% of the data) is used for RF model training, and the same 15\% test set is used for evaluation.

\subsection{Data with simulated noise}
\label{sec: Simulation}

We simulated data based on the ChEMBL data sets with both homoscedastic and heteroscedastic variance to better evaluate the prediction interval results under known distributions.
In this exercise we added artificial noise to the observed activity. We would expect the prediction intervals to reflect whatever noise we artificially added. 
For each ChEMBL dataset, we built an RF model and use the out-of-bag predictions $\hat{y} = f(X)$ as the true activity of a molecule with descriptor $X$. 
Then we simulated the measured molecular activity $\tilde{y}$ with homoscedastic or heteroscedastic noise $e \sim \mathbf{N}(0, \sigma^2)$: $\tilde{y} = f(X) + e$. 

\begin{itemize}
    \item Case 1: Homoscedastic variance: $\sigma$ is a constant in each dataset. 
    \item Case 2: Heteroscedastic variance: $\sigma$ is an S-shape monotone increasing function of $f(X)$, with the average value the same as the constant standard deviation of the homoscedastic error in Case 1 for the same dataset.
\end{itemize}

\begin{figure}[htbp]
\caption{Real vs. simulated activities of the Estrogen dataset. Top panel: the standard deviation $\sigma$ of added noise versus the true molecular activity f(X) in both simulation cases, where the black curve is the $\sigma$ in Case 2 and the gold horizontal line is the $\sigma$ in Case 1. Bottom panel: scatter plots of the new $\tilde{y}$ versus true activity f(X) in two cases; and the red dash line is the identity line.}
\centering
\includegraphics[width=0.8\textwidth]{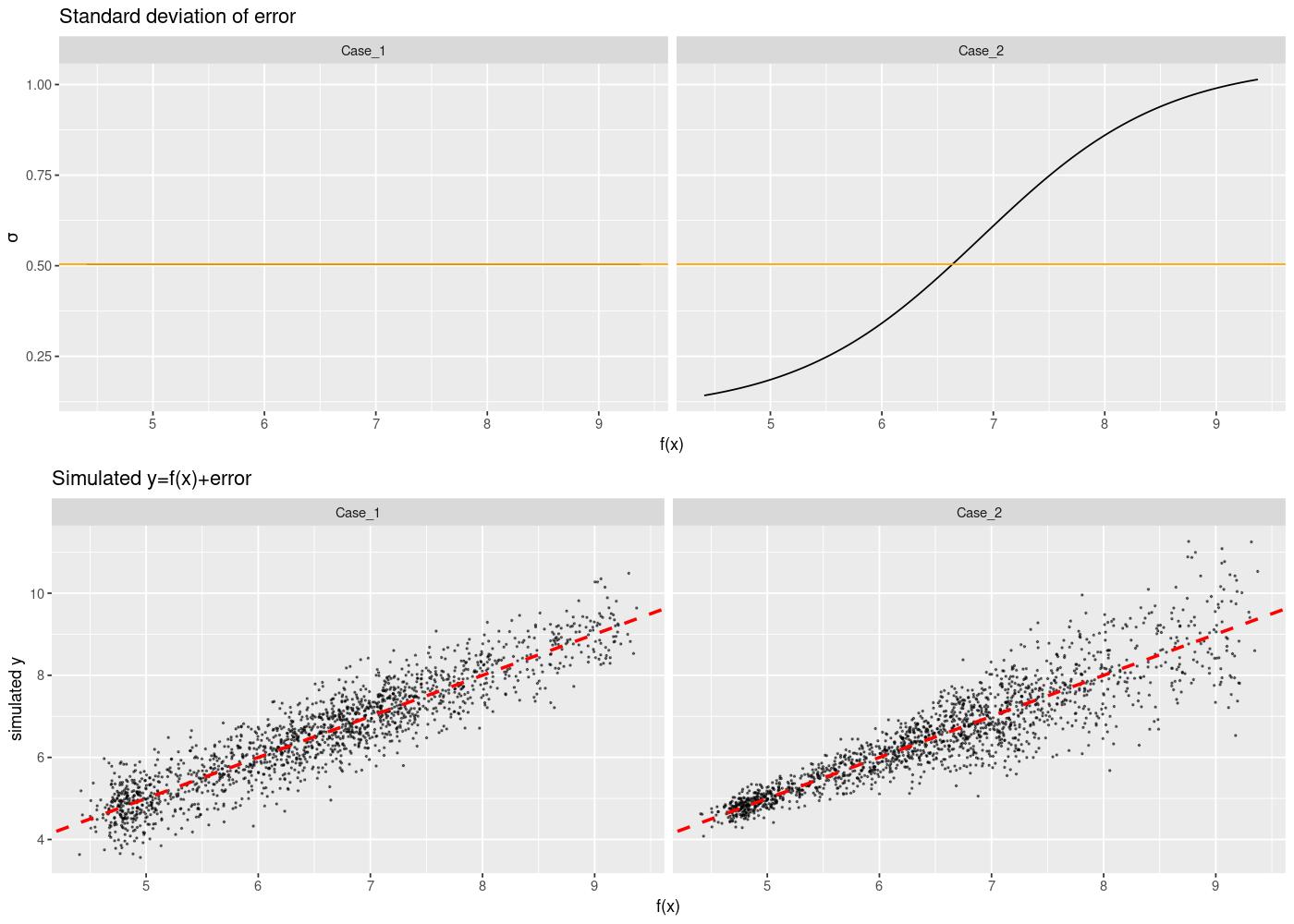}
\label{fig:Simulation}
\end{figure}

Use the simulated data based on the Estrogen dataset as an example: Figure \ref{fig:Simulation} shows the relationships between $\sigma$ and $f(X)$, as well as the scatter plots of simulated $\tilde{y}$ against the truth $f(X)$ in both simulation cases. The details of the simulation settings are provided in Supporting Information Section \ref{SI:SimulationDetails}.

Table \ref{table:simulation_summary} provides a summary of the average predictive performance and computational time of each machine learning prediction model for both simulation studies. The average is over all datasets. All models achieved a satisfactory prediction accuracy with R-squared around 0.6. Since the underlying true relationships were fitted using RF model, this models achieved the best performance. 
\begin{table}[htbp]
\caption{Summary of average predictive performance on test sets and run time for each ML algorithm in simulation studies.}
\label{table:simulation_summary}
\centering
\resizebox{\columnwidth}{!}{
\begin{tabular}{|l|lll|lll|}
\hline
                                 & \multicolumn{3}{c|}{\textbf{Simulation 1}}                                & \multicolumn{3}{c|}{\textbf{Simulation 2}}                                \\ \hline
\textbf{ML Prediction Algorithm} & \multicolumn{1}{l|}{R-squared} & \multicolumn{1}{l|}{RMSE}  & Runtime (s) & \multicolumn{1}{l|}{R-squared} & \multicolumn{1}{l|}{RMSE}  & Runtime (s) \\ \hline
RF                               & \multicolumn{1}{l|}{0.664}     & \multicolumn{1}{l|}{0.621} & 43.5        & \multicolumn{1}{l|}{0.626}     & \multicolumn{1}{l|}{0.673} & 44.5        \\ \hline
DNN-dropout                      & \multicolumn{1}{l|}{0.613}     & \multicolumn{1}{l|}{0.666} & 80.3        & \multicolumn{1}{l|}{0.565}     & \multicolumn{1}{l|}{0.730} & 83.1        \\ \hline
DNN-multitask                    & \multicolumn{1}{l|}{0.638}     & \multicolumn{1}{l|}{0.646} & 25.6        & \multicolumn{1}{l|}{0.595}     & \multicolumn{1}{l|}{0.706} & 23.9        \\ \hline
GB                               & \multicolumn{1}{l|}{0.645}     & \multicolumn{1}{l|}{0.636} & 8.07        & \multicolumn{1}{l|}{0.605}     & \multicolumn{1}{l|}{0.693} & 7.86        \\ \hline
\end{tabular}
}
\end{table}

\begin{figure}[htbp]
\centering
\caption{Box plots of marginal coverage errors in simulations for nominal coverage 70\% (left), 80\% (middle) and 90\% (right). Top row: Simulation Case 1; bottom row: Simulation Case 2.}
\includegraphics[width=\textwidth]{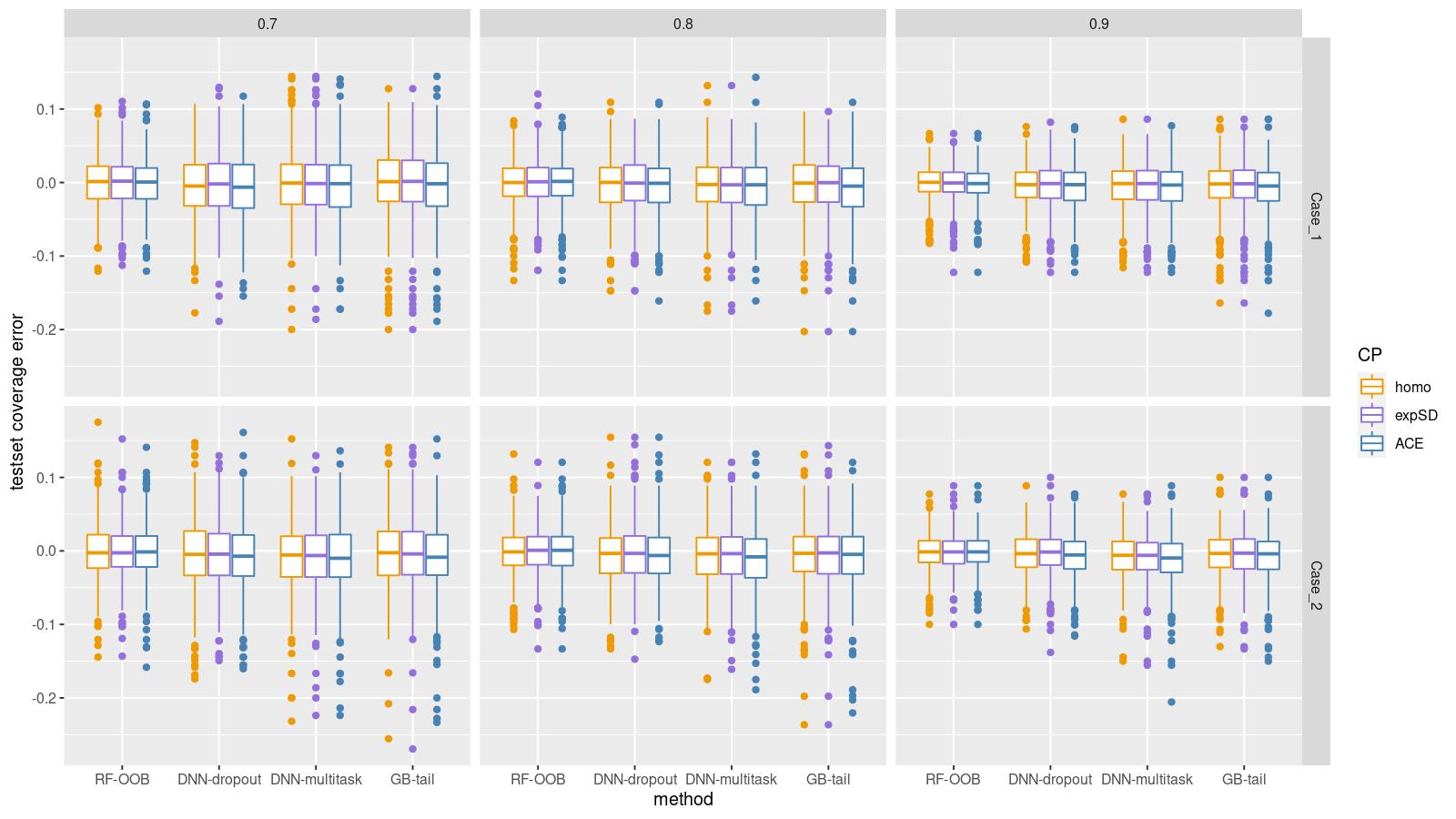}
\label{fig:sim_coverageMarginal}
\end{figure}

\begin{figure}[htbp]
\centering
\caption{Average no-model normalized Prediction Interval widths in simulations for nominal coverage 70\% (left), 80\% (middle) and 90\% (right). Top row: Simulation Case 1; bottom row: Simulation Case 2. 
}
\includegraphics[width=\textwidth]{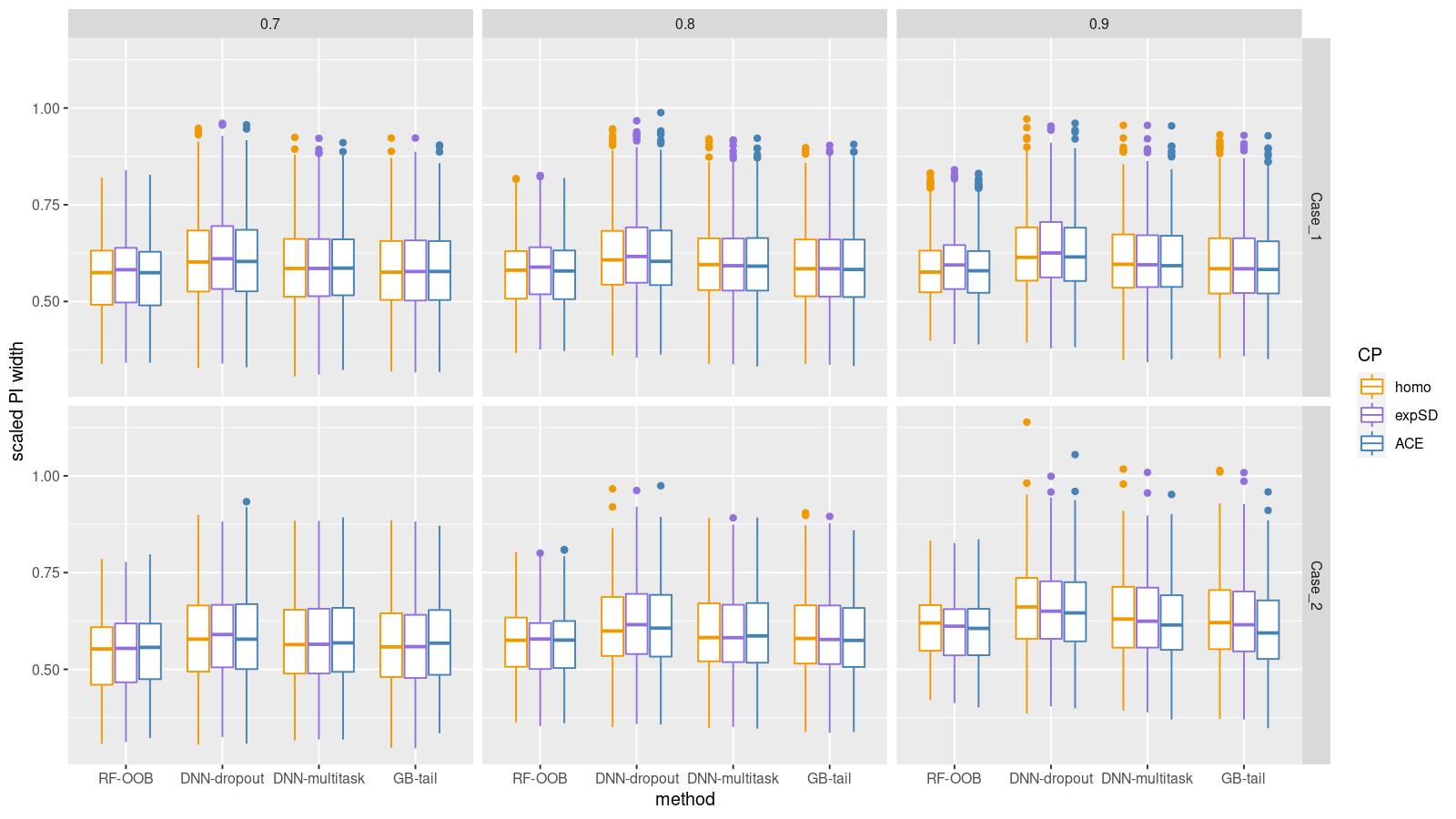}
\label{fig:sim_PIwidth}
\end{figure}

We created conformal predictors using three CP algorithms (homo, expSD, ACE) and four ML methods (RF-OOB, DNN-dropout, DNN-multitask, and GB) under three nominal coverage levels: 70\%, 80\% and 90\%. 
Figure \ref{fig:sim_coverageMarginal} shows the box plots of test set coverage errors. These are centered around zero in all scenarios, as expected since conformal prediction intervals have valid marginal coverage. 
Figure \ref{fig:sim_PIwidth} evaluates the efficiency of conformal predictors using the average prediction interval widths. The prediction intervals are scaled by the ``no-model'' case interval width to be comparable across datasets. Overall, the average sizes of prediction intervals are similar across different CP algorithms and/or ML methods.

\begin{figure}[htbp]
\centering
\caption{Test set coverage of subgroups defined by prediction interval widths, at nominal level 80\% (the horizontal dash line). 
Top row: Simulation case 1; bottom row: Simulation case 2.
The five subgroups in each test set were created by sorting the prediction interval widths in increasing order and dividing the molecules into five equal size bins; so that, e.g., subgroup 1 contains the molecules with the shortest prediction interval widths. 
}
\includegraphics[width=\textwidth]{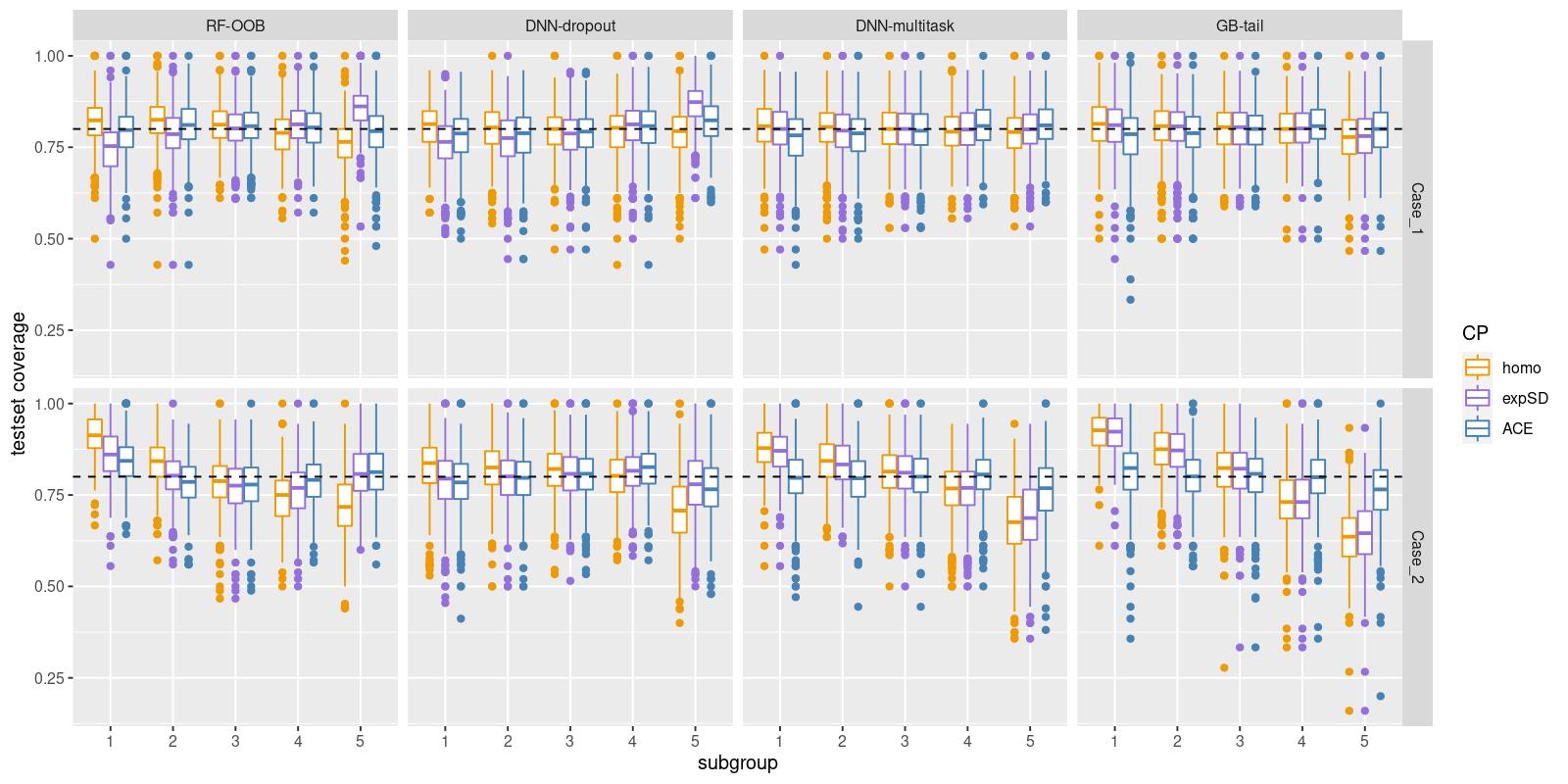}
\label{fig:sim_subGroup_80}
\end{figure}

\begin{figure}[htbp]
\centering
\caption{Average no-model normalized prediction interval widths of subgroups defined by prediction interval widths, at nominal level 80\%. 
Top row: Simulation case 1; bottom row: Simulation case 2. 
The five subgroups in each test set were created by sorting the prediction interval widths in increasing order and dividing the molecules into five equal size bins; so that, e.g., subgroup 1 contains the molecules with the shortest prediction interval widths. 
}
\includegraphics[width=\textwidth]{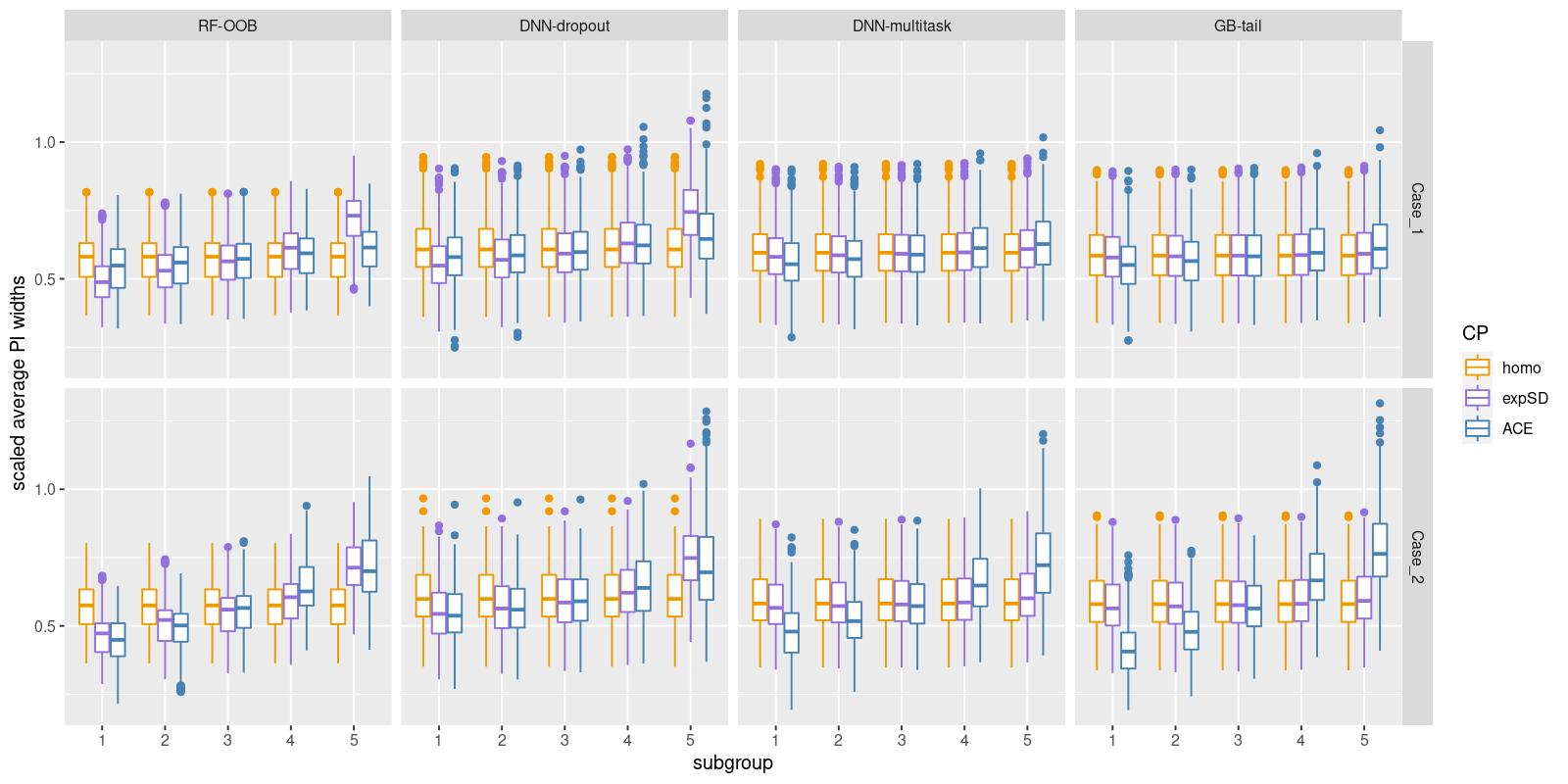}
\label{fig:sim_subGroup_80_avgPIwidth}
\end{figure}

To demonstrate conditional coverage properties of different conformal predictors, for each ML prediction model and each data set sorted the test set molecules by their ML method-specific prediction interval widths, from small to large, and then split them into five equal-size sub-groups by binning these widths. We then plotted box plots of the prediction interval coverage (Figure \ref{fig:sim_subGroup_80}) and width (Figure \ref{fig:sim_subGroup_80_avgPIwidth}) within each bin. In simulation Case 1, when the variance of errors is constant, it is more efficient to produce the constant-width prediction interval. This is the case with the ACE CP algorithm. The expSD CP algorithm performs well for DNN-multitask and GB-tail, but often undercovers the nominal levels in the first and overcovers them in the fifth subgroup for RF-OOB and DNN-dropout methods.
In Case 2, when the data is simulated with large heteroscedastic variance, the homo intervals severely overcover the nominal levels in the first two subgroups corresponding to a relatively small prediction uncertainty while
undercover the nominal levels in the last three subgroups. The ACE algorithm successfully provides adaptive prediction interval widths, so that the test set coverages in all subgroups are closer to the nominal 0.8 indicated by the dashed line. DNN-multitask and GB-tail algorithms perform similarly to homo CP in both coverage and average prediction interval widths. The expSD algorithm, on the other hand, performs better than homo CP for RF-OOB and DNN-dropout methods, however, for DNN-multitask and GB-tails its performance reduces to that of homo CP. Similar results for conditional coverage comparisons indicating competitiveness of the proposed ACE CP algorithm were observed for nominal levels 70\% and 90\%, which are included in the Supporting Information Section \ref{SI:AdditionalFigures}.

\begin{figure}[htbp]
\centering
\caption{Prediction uncertainty scores, s(X), from different datasets and different ML methods (Simulation Case 2).}
\includegraphics[width=\textwidth]{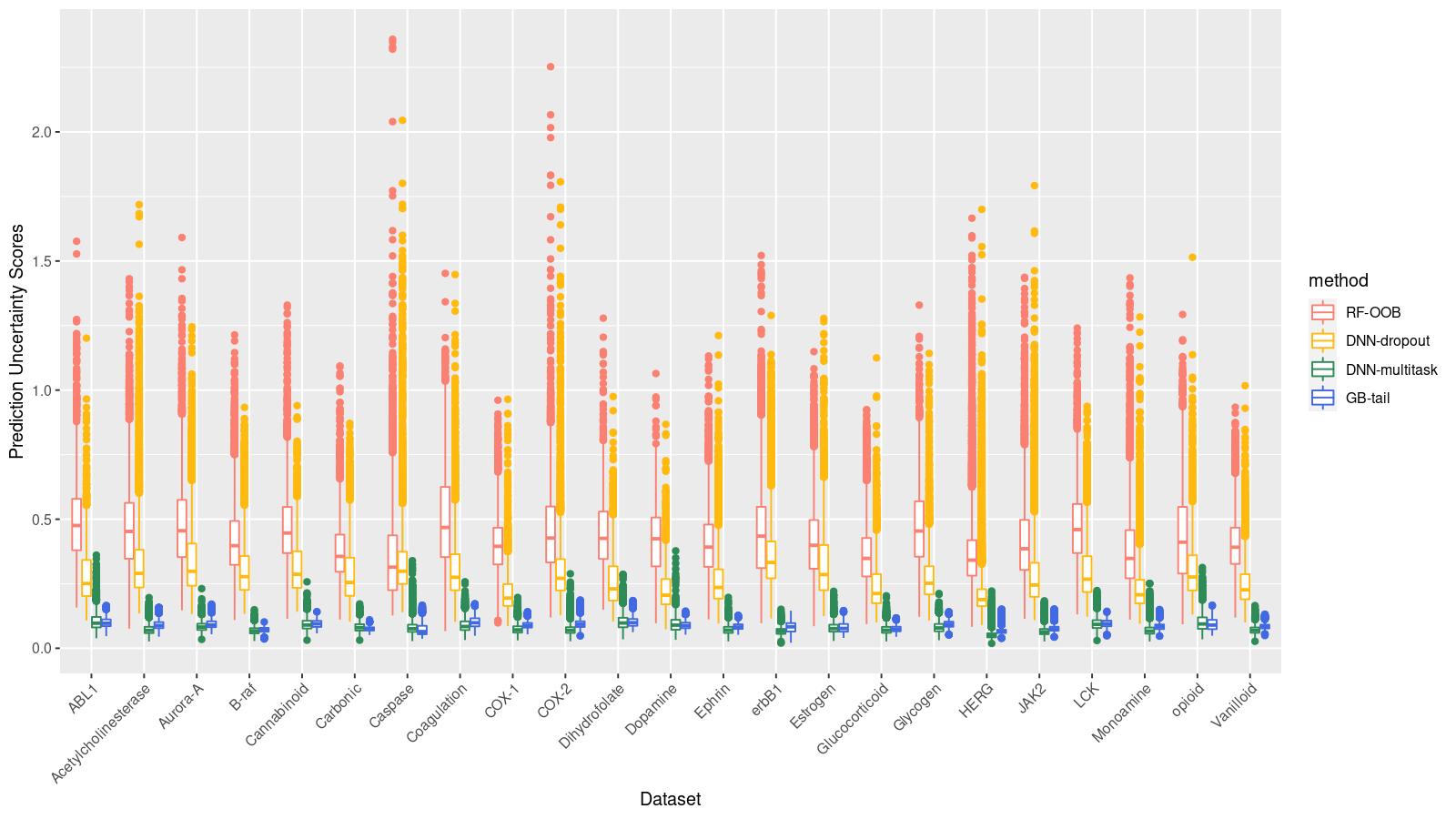}
\label{fig:sim2_PredSD}
\end{figure}

\begin{figure}[htbp]
\centering
\caption{Prediction interval widths comparison for nominal coverage 90\% (Simulation Case 2). The prediction intervals are normalized by ``no-model'' interval width. The horizontal dash line at 1.0 is the scaled ``no-model'' interval width.}
\includegraphics[width=\textwidth]{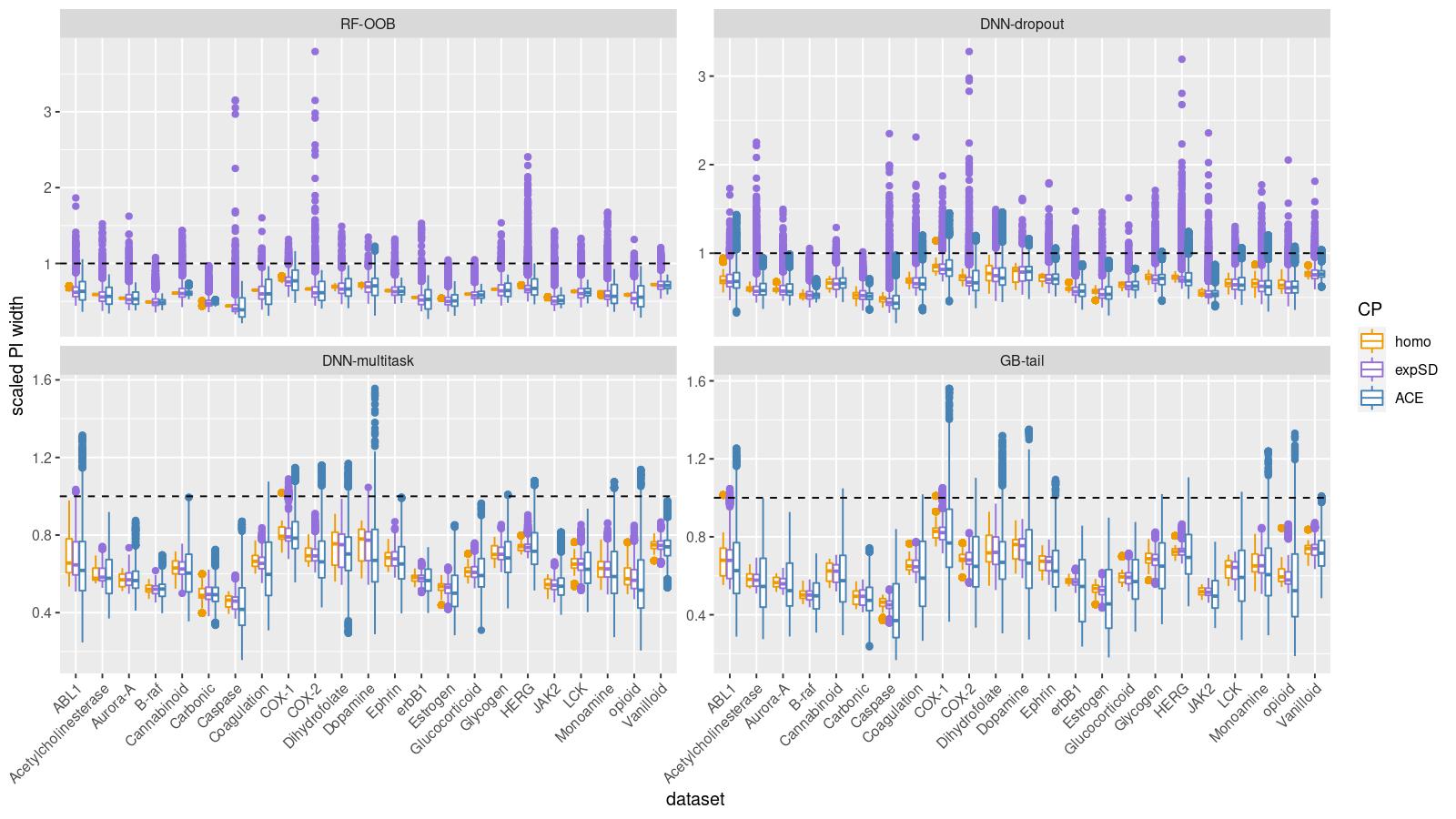}
\label{fig:sim2_PIwidth_dataset}
\end{figure}

To better understand the performance variability of expSD when it is applied to different ML methods, we compared the raw prediction uncertainty scores from different ML methods in Figure \ref{fig:sim2_PredSD}, using Simulation Case 2 results as an example. The RF-OOB and DNN-dropout methods always produced raw prediction uncertainty scores with large variabilities within a dataset. In contrast, the raw prediction uncertainty scores from DNN-multitask and GB-tail have a much smaller range and are very close to zero, which leads to a close to constant scaling factor in expSD and thus similar to homo prediction intervals in both simulation cases. For RF-OOB and DNN-dropout methods, the non-adaptive expSD transformation created heteroscedastic prediction intervals even when the errors are homoscedastic, and frequently produced very wide intervals for some molecules, especially at high nominal coverage levels. Figure \ref{fig:sim2_PIwidth_dataset} shows the boxplots of prediction interval widths at a nominal level of 90\%. In many datasets, there are molecules with expSD intervals that are twice as wide as the ``no-model'' interval width, which is not compatible with our expectations. 
We thus decided to exclude the expSD CP algorithm from the application section and future comparisons. 

\subsection{Application to ChEMBL and Kaggle data sets}
\label{sec:Application}

The average predictive performance and computational time over repeats for two groups of datasets are listed in Table \ref{table:application_summary}. All models achieved satisfactory prediction accuracy. The two DNN models and the GB model perform slightly better than RF, which usually used as the baseline model in QSAR modeling, despite that the RF model uses more training data (85\% of an entire dataset) compared to others (70\% of an entire dataset). Also, the DNN-multitask and GB models took less time compared to DNN-dropout and RF. Although the point predictor accuracy and run time may not affect the performance of conformal predictors, these comparisons demonstrated the need of developing suitable uncertainty quantification methods tailored for all kinds of practical QSAR models, in additional to the widely used RF-based conformal predictors.

\begin{table}[htbp]
\caption{Summary of test set predictive performance and run time (in seconds) of each ML method for ChEMBL and Kaggle datasets. These numbers represent an average over the datasets.}
\label{table:application_summary}
\resizebox{\columnwidth}{!}{
\begin{tabular}{|l|lll|lll|}
\hline
                                 & \multicolumn{3}{c|}{\textbf{ChEMBL}}                                      & \multicolumn{3}{c|}{\textbf{Kaggle}}                                      \\ \hline
\textbf{ML Prediction Algorithm} & \multicolumn{1}{l|}{R-squared} & \multicolumn{1}{l|}{RMSE}  & Runtime (s) & \multicolumn{1}{l|}{R-squared} & \multicolumn{1}{l|}{RMSE}  & Runtime (s) \\ \hline
RF                               & \multicolumn{1}{l|}{0.633}     & \multicolumn{1}{l|}{0.710} & 41.0        & \multicolumn{1}{l|}{0.683}     & \multicolumn{1}{l|}{0.565} & 1314.3      \\ \hline
DNN-dropout                      & \multicolumn{1}{l|}{0.652}     & \multicolumn{1}{l|}{0.688} & 127.8       & \multicolumn{1}{l|}{0.703}     & \multicolumn{1}{l|}{0.541} & 893.4       \\ \hline
DNN-multitask                    & \multicolumn{1}{l|}{0.645}     & \multicolumn{1}{l|}{0.697} & 36.2        & \multicolumn{1}{l|}{0.700}     & \multicolumn{1}{l|}{0.547} & 781.3       \\ \hline
GB                               & \multicolumn{1}{l|}{0.644}     & \multicolumn{1}{l|}{0.694} & 8.4         & \multicolumn{1}{l|}{0.706}     & \multicolumn{1}{l|}{0.536} & 143.8       \\ \hline
\end{tabular}
}
\end{table}

\begin{figure}[htbp]
\centering
\caption{Boxplots of marginal coverage errors in two groups of datasets for eight nominal coverages ranging from 60\% to 95\%. Top row: ChEMBL datasets; bottom row: Kaggle datasets.}
\includegraphics[width=\textwidth]{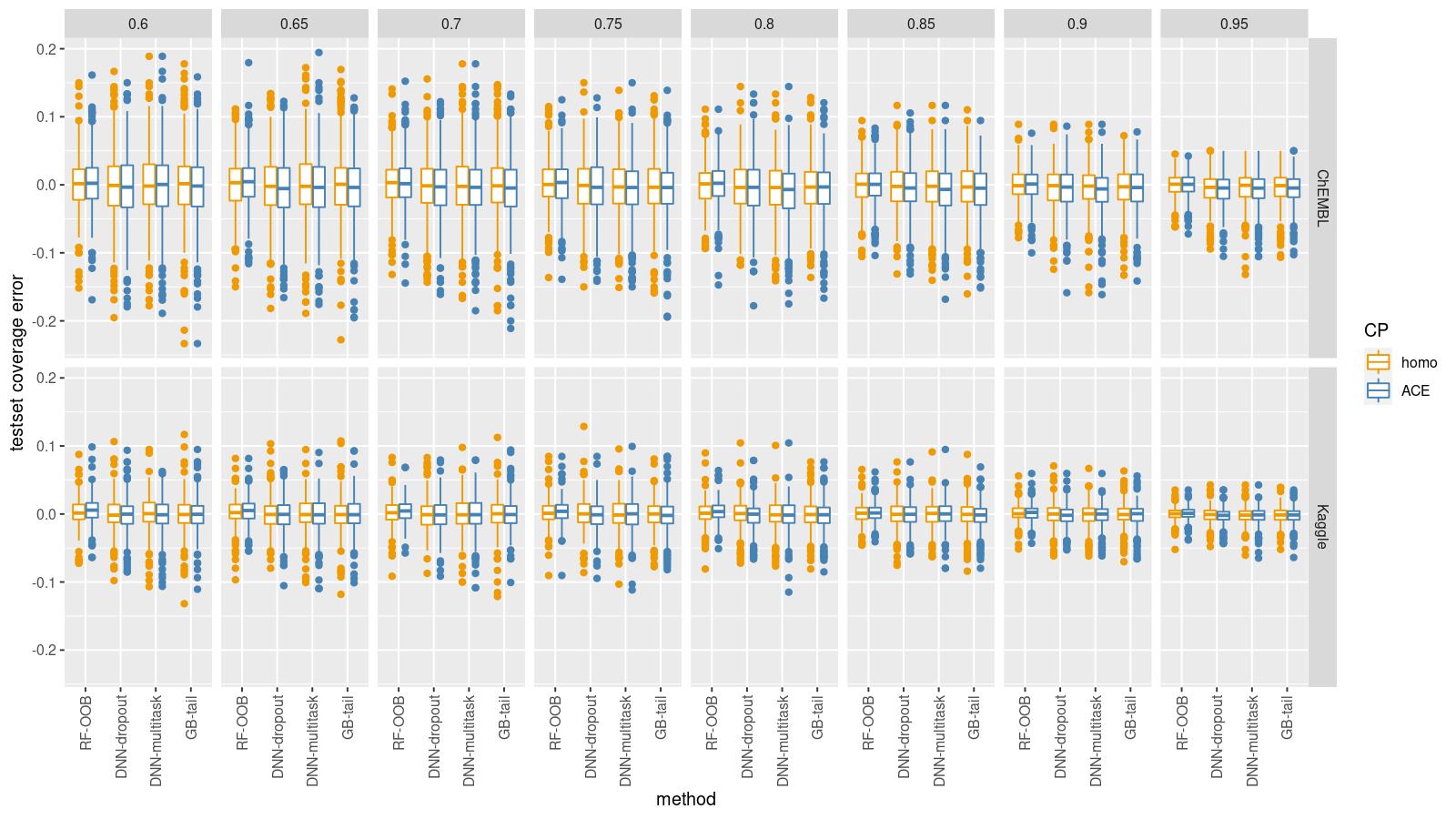}
\label{fig:application_coverageMarginal_boxplot}
\end{figure}

\begin{figure}[htbp]
\centering
\caption{The absolute marginal coverage errors averaged across datasets and repeats, for eight nominal coverages ranging from 60\% to 95\% and two groups of datasets. Top row: ChEMBL datasets; bottom row: Kaggle datasets.}
\includegraphics[width=0.8\textwidth]{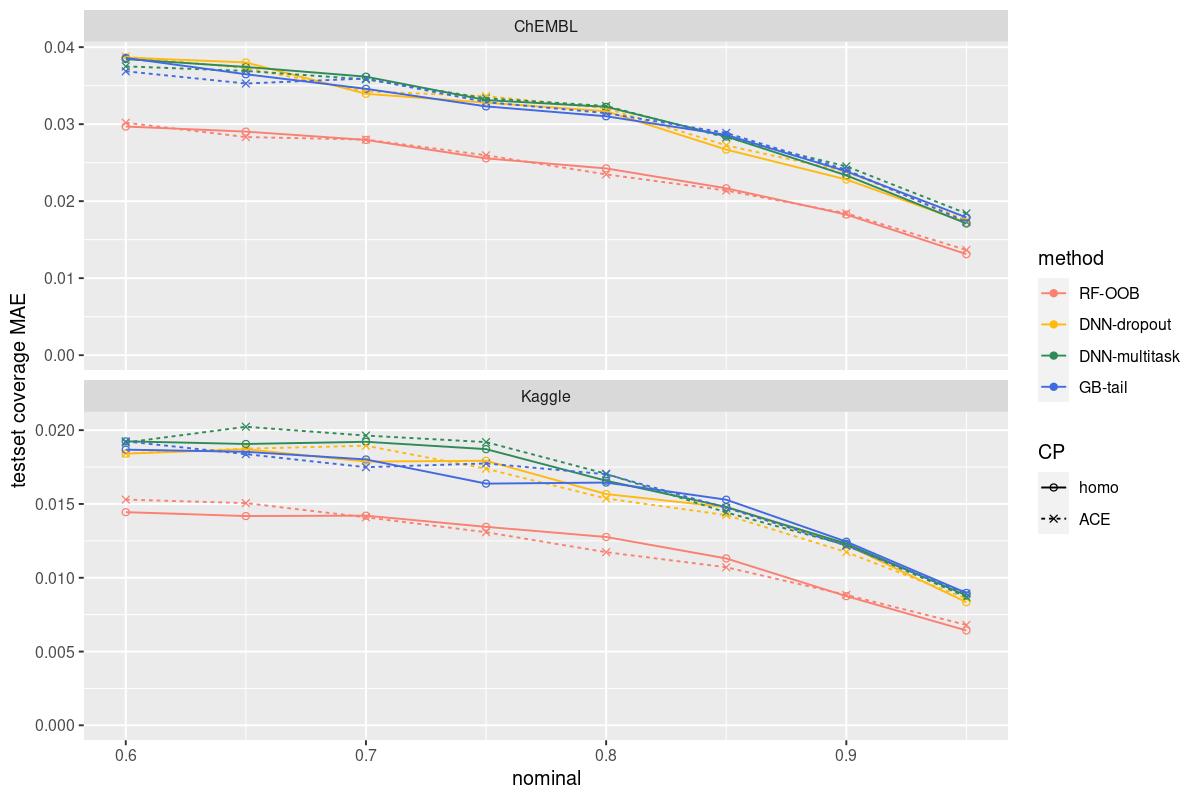}
\label{fig:application_coverageMarginal_MAE}
\end{figure}

We created conformal predictors using two CP algorithms (homo, ACE) and four ML methods (RF-OOB, DNN-dropout, DNN-multitask, GB) under eight nominal coverage levels ranging from 60\% to 95\% with increments of 5\%.  
Figure \ref{fig:application_coverageMarginal_boxplot} shows the boxplots of marginal coverage error on test sets under different nominal coverage levels for each method and CP algorithm. In all cases the coverage errors are centered around zero, which indicates that they all achieved marginal validity. Since the Kaggle datasets are larger than ChEMBL datasets, the distributions of errors have less variability. 
The comparison of variability in marginal coverage errors between ML method and CP algorithm is shown in Figure \ref{fig:application_coverageMarginal_MAE}: For each ML method and CP algorithm, we averaged the absolute value of coverage errors across datasets and repeats at multiple nominal coverages. The RF-OOB method has lower mean absolute errors of marginal coverage in all nominal levels. The RF model is trained on all the available data without splitting into proper training set and calibration set, and it used the out-of-bag data for calibration, which is effectively larger than the stand alone calibration sets in other methods.   Due to this unique feature in RF, it achieved more robust performance in the marginal coverage. There is no difference between ACE and homo, which is expected.

\begin{figure}[htbp]
\centering
\caption{Boxplots of average scaled prediction interval widths for eight nominal coverages ranging from 60\% to 95\%. }
\begin{subfigure}[b]{\textwidth}
\includegraphics[width=\textwidth]{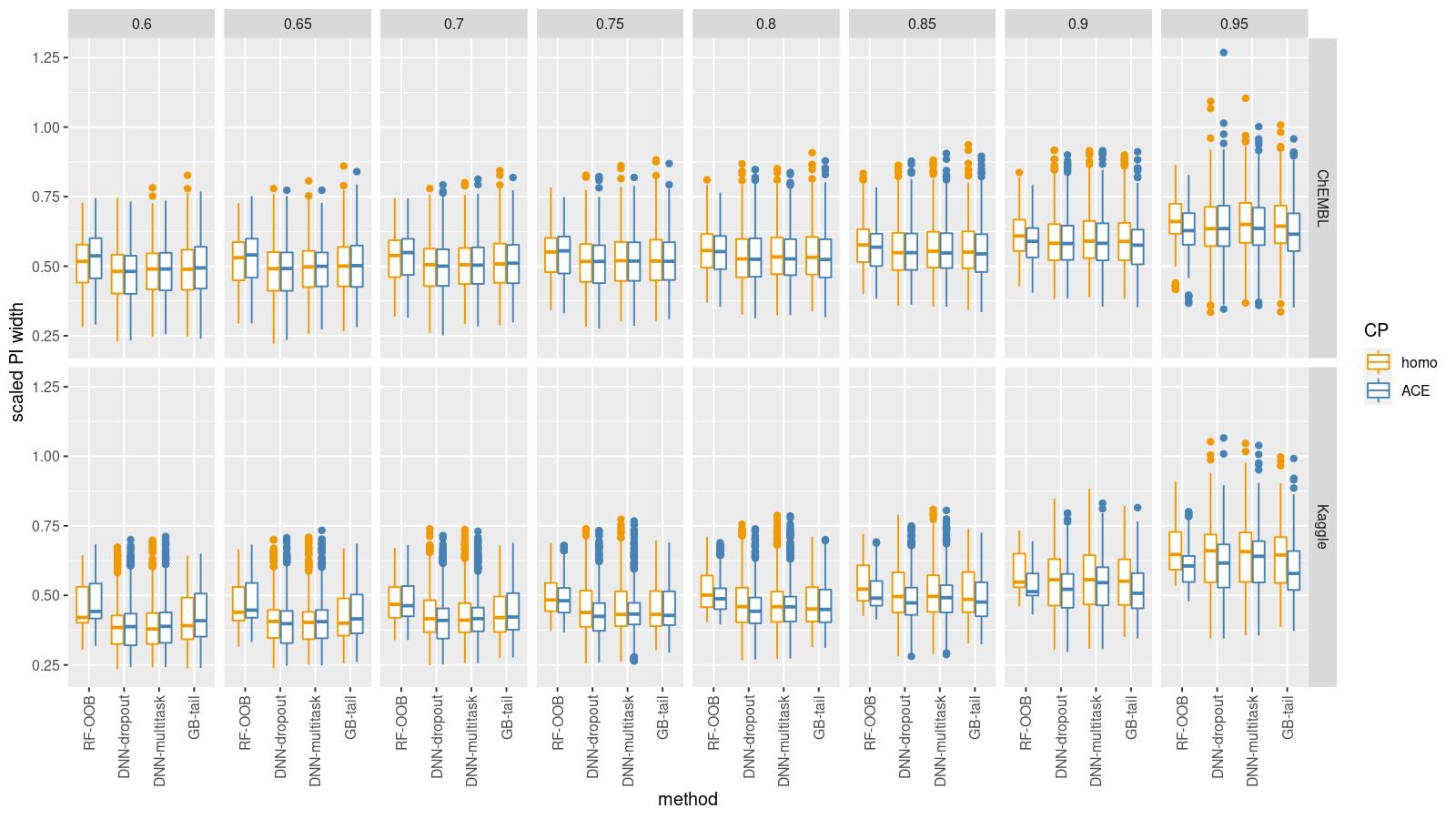}
\caption{Scaled the prediction Intervals by ``no-model'' interval width. Top row: ChEMBL datasets; bottom row: Kaggle datasets.}
\end{subfigure}
\begin{subfigure}[b]{\textwidth}
\includegraphics[width=\textwidth]{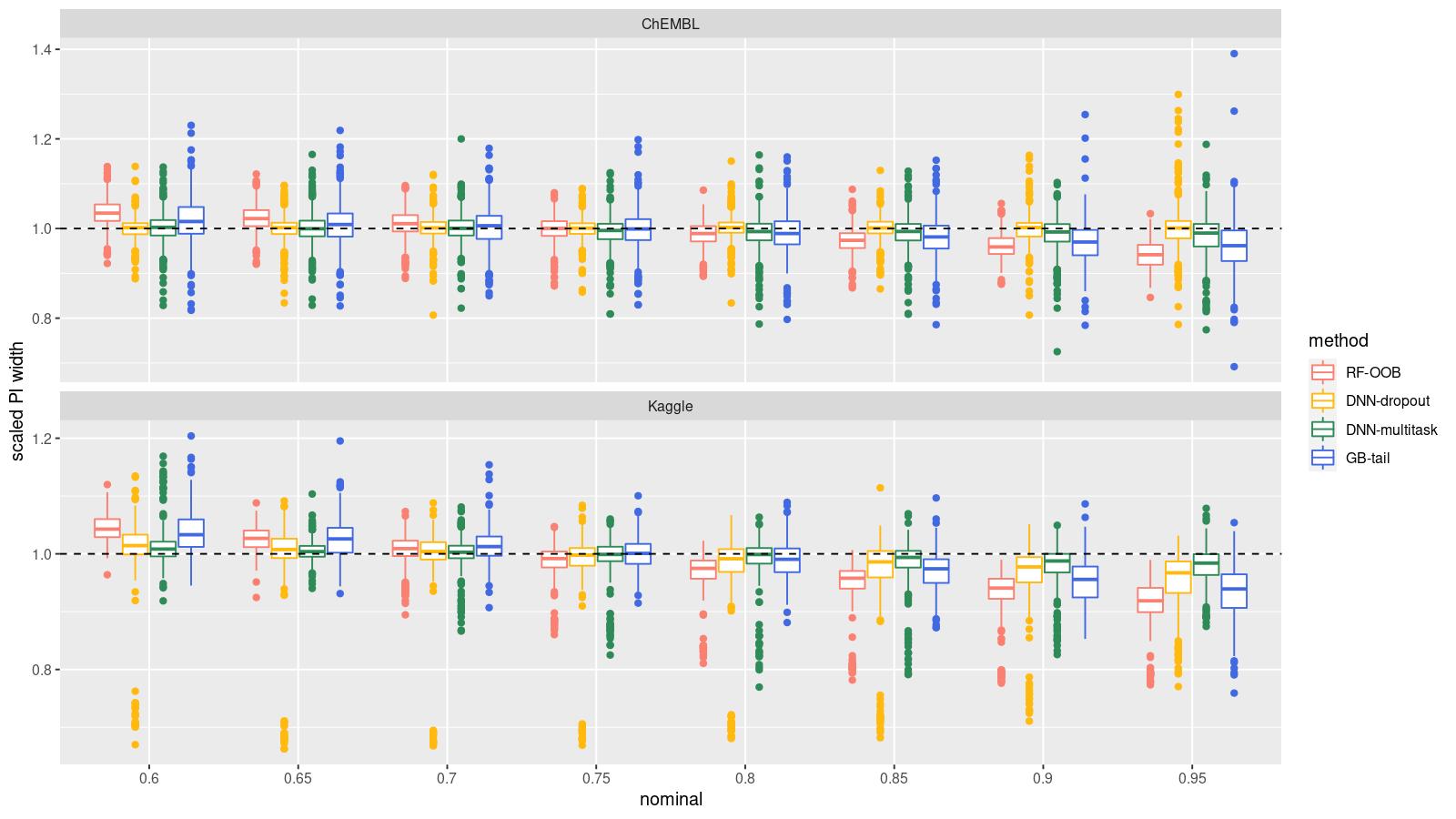}
\caption{Scaled the prediction Intervals by homo interval width. Top row: ChEMBL datasets; bottom row: Kaggle datasets.}
\end{subfigure}
\label{fig:application_PIwidth}
\end{figure}

Figure \ref{fig:application_PIwidth} compares the efficiency of conformal predictors. 
In the top subplot, the DNN-dropout, DNN-multitask, and GB-tail methods have slightly narrower average prediction interval widths than the RF-OOB method in the lower nominal range (60\% to 85\%), but the distributions of their average prediction interval widths have a wider spread. Also, there are a few outliers at the higher nominal level of 95\%. 
The bottom subplot compares the efficiency between ACE and homo conformal algorithms. For each ML method, the average prediction interval width from the ACE algorithm of each test set is scaled by the corresponding homo prediction interval width. As the nominal level increases, the ACE algorithm is more likely to produce narrower prediction intervals, on average, compared to homo. 

\begin{figure}[htbp]
\centering
\caption{Test set coverage of subgroups defined by prediction interval widths, at nominal level 80\%}
\includegraphics[width=\textwidth]{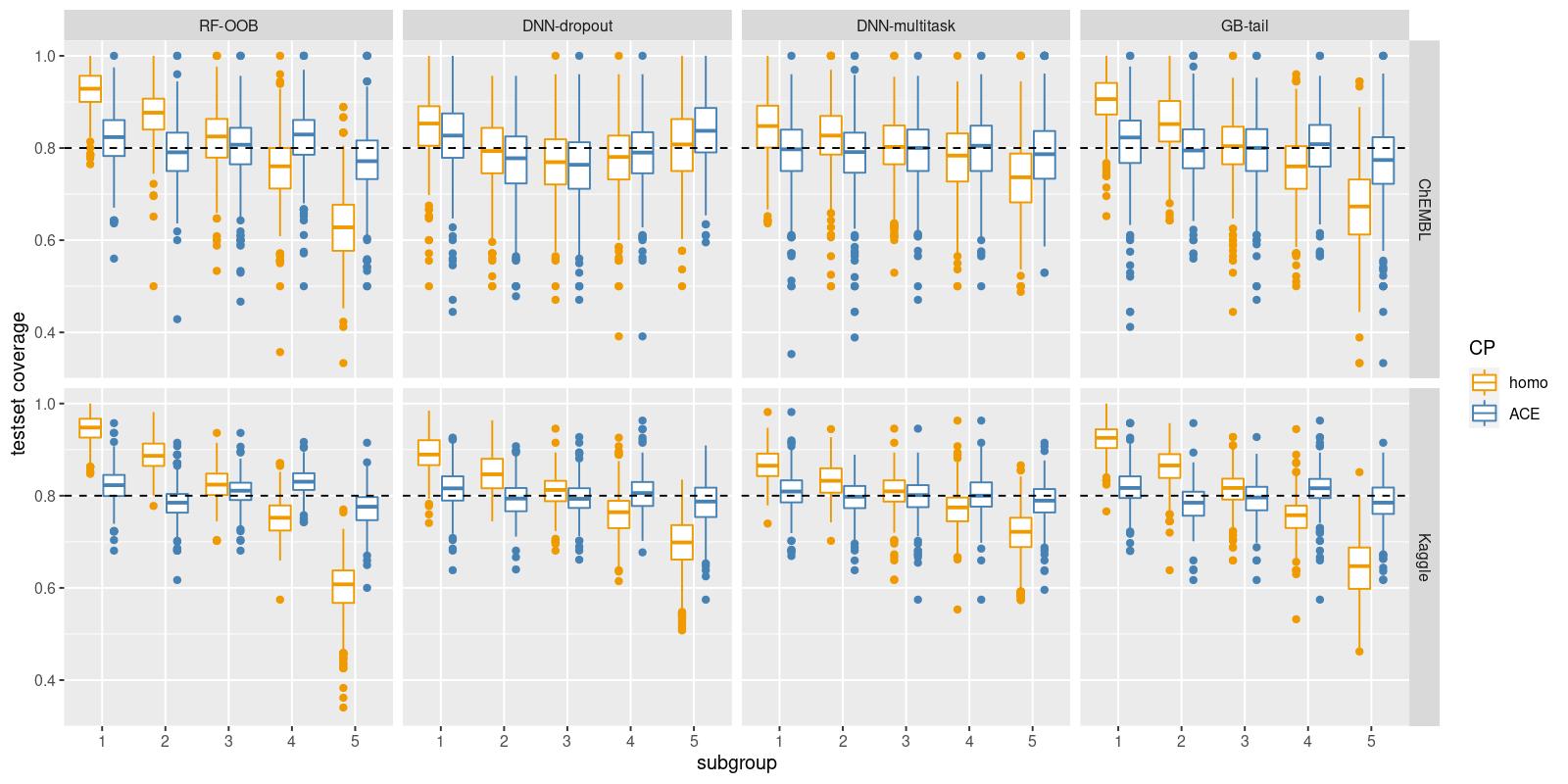}
\label{fig:application_subGroup_PIwidth_80}
\end{figure}

\begin{figure}[htbp]
\centering
\caption{The absolute error of test set coverage by ACE algorithm, averaged across subgroups.}
\includegraphics[width=\textwidth]{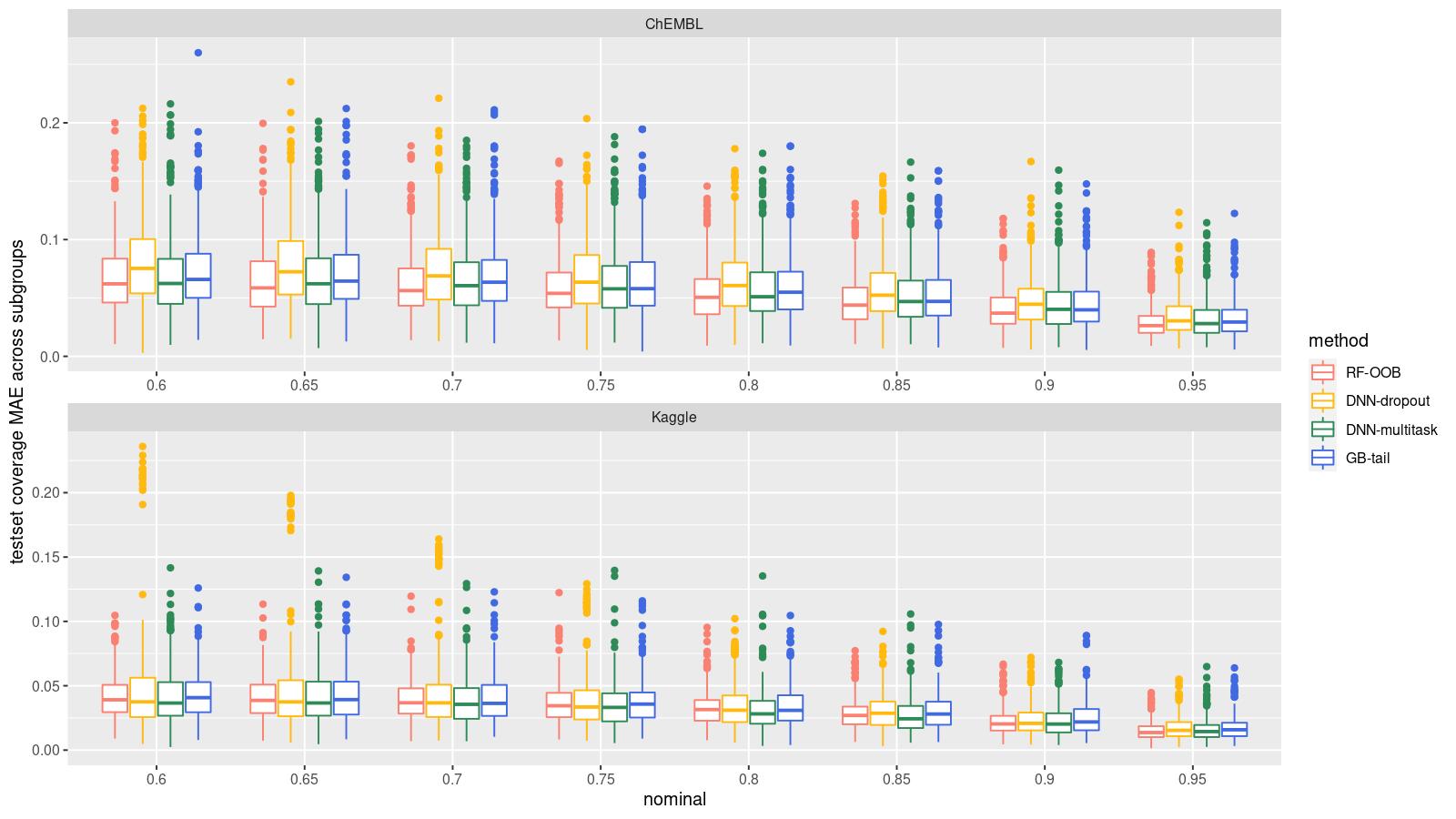}
\label{fig:application_subGroup_PIwidth_ACE}
\end{figure}

Figure \ref{fig:application_subGroup_PIwidth_80} demonstrates the conditional coverage property of ACE in contrast to homo using the nominal level of 80\% as an example. Again, each test set was sorted by the prediction interval widths for each method in increasing order, and split into five equal-size subgroups. 
Similar to what we observed in the simulation study, the ACE algorithm achieves closer to the expected nominal coverage (dashed line) in all subgroups compared to homo. 
In most situations, except the DNN-dropout model for ChEMBL datasets, there is a clear decreasing trend of homo prediction interval coverage from the first subgroup to the fifth subgroup. It indicates that most of the molecules with smaller prediction intervals have smaller prediction errors, which lead to over-coverage of homo PIs in the first two subgroups. Also, the molecules in the last two subgroups have larger prediction errors than the PI width and hence this results in under coverage of homo PIs. In other words, the decreasing trend of homo PI coverage demonstrates that the prediction interval effectively differentiates between the accurately predicted molecules from those with large prediction errors. Thus the prediction intervals from DNN-dropout ensembles in some cases are not as informative as the other methods. 
Figure \ref{fig:application_subGroup_PIwidth_ACE} compares the conditional coverage of ACE prediction intervals generated by different methods. For ChEMBL datasets, the DNN-dropout model shows a higher mean absolute error of coverage by subgroups at multiple nominal levels. For some Kaggle datasets, the DNN-dropout model also produced noticeably higher errors at lower nominal levels. 

\begin{figure}[htbp]
\centering
\caption{
Association of the ACE prediction interval widths and absolute prediction errors. (Nominal level of 80\% for the Kaggle dataset 3A4. Shown results are for one test set out of 20 repeated runs). 
The orange line is the prediction interval width from ACE algorithm. Each dot in the scatter plot is a test set molecule, colored by whether it is successfully covered by the prediction interval (with absolute error below the orange line). Both the absolute errors and the prediction interval width are scaled by the corresponding homo prediction interval width. The horizontal black dash line with y-intercept at 1.0 represents the scaled homo prediction interval size (unit length). }
\includegraphics[width=\textwidth]{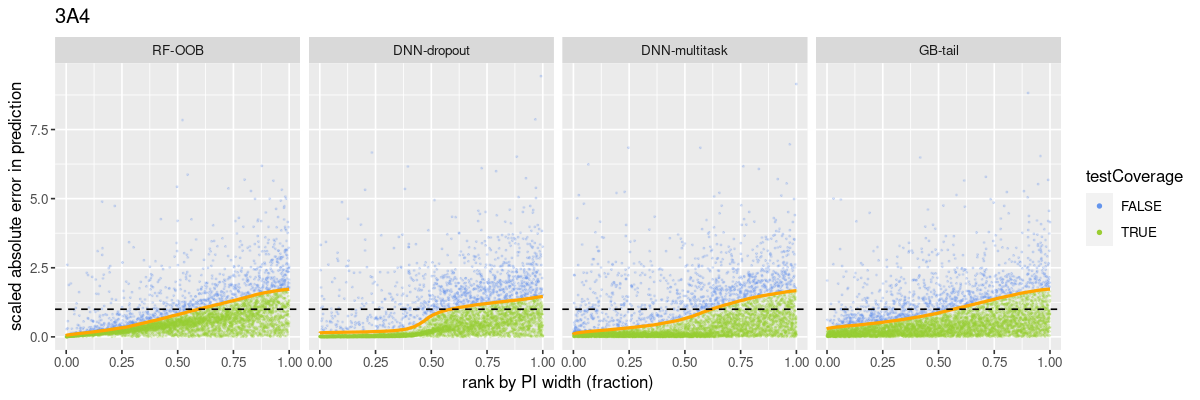}
\label{fig:application_absErr-PIrank}
\end{figure}

Figure \ref{fig:application_absErr-PIrank} visualizes the ACE prediction intervals using the Kaggle 3A4 dataset as an example. The molecular activities in the 3A4 dataset have a truncated distribution causing heteroscedastic prediction errors. In each subplot, the ranking of test set molecules on the horizontal axis is determined by the prediction interval widths from the corresponding ML method. The orange line shows that the spread of the absolute errors is increasing with the prediction interval width for all the four methods, indicating a strong association between the magnitude of true prediction error and prediction interval width.
Thus, with the ACE algorithm, we obtain well-calibrated heteroscedastic prediction intervals with the widths adaptive to the true absolute prediction errors. 
Empirical coverage of the ACE prediction intervals reflected by the proportion of points below the regression line (numbers are not shown) is close to the nominal levels at each interval width, Thus, the ACE method provides meaningful estimates of the prediction errors. 

\section{Discussion}
\label{sec: Discussion}

In this work, we developed the ACE algorithm, an inductive conformal predictor whose nonconformity scores are calculated using estimates of the prediction uncertainty generated by the ensemble of point prediction models. We evaluated this method using both simulated and large collections of real QSAR data. The ACE algorithm produced prediction intervals that are strongly adapted to data: for homoscedastic data, the interval width is nearly constant; for heteroscedastic data, the width varies with the magnitude of prediction errors and achieves close to nominal coverage for each molecule.
This adaptiveness feature of ACE prediction intervals makes it informative in QSAR applications. One can use the interval size to differentiate the molecules which are accurately predicted vs. those that are not.

The ACE algorithm is applicable to any prediction model, if one can compute a raw prediction uncertainty score correlated with the variance of prediction error.
Some of the widely used QSAR methods, such as DNNs, provide highly accurate point predictions but do not generate raw uncertainty scores because they do not naturally produce ensembles. The GB method, although an ensemble method but unlike RF does not generate a prediction uncertainty estimate either. We proposed the novel ensemble DNN-multitask and GB-tail methods, to compute the raw prediction uncertainty score from DNN and GB point prediction models with a little extra computational cost. The ACE prediction intervals with raw prediction uncertainty scores obtained by both these methods showed competitive performance in a diverse collection of QSAR datasets. For neural net models, we see that using DNN-multitask produces more useful ensembles compared to DNN-dropout. 
Implementing the DNN-multitask method involves only a simple re-formatting of the training molecular activities without changes in the neural net structure. This method could be also applied to the convolutional neural network\cite{lecun1998gradient,meyer2019learning,shi2019molecular} or graph convolutional networks\cite{duvenaud2015convolutional,kearnes2016molecular,wu2018moleculenet,yang2019analyzing}.

In conclusion, the conformal predictors computed by the proposed ACE algorithm jointly with highly accurate and commonly used ML models may serve as practical uncertainty quantification tools for QSAR modeling, producing accurate and informative prediction intervals without compromising the point prediction quality or computational efficiency. 


\newpage
\section*{Supporting Information}

\subsection{ACE pseudo code}
\label{SI:pseudocode}

\begin{algorithm}[H]
	\caption{Adaptive Calibrated Ensemble (ACE)}
	\label{alg:ACE}
	\begin{algorithmic}[1]
	    \State Calculate $\tilde{s} = \frac{s - \mu_s}{sd_s}$, where $\mu_s$ and $sd_s$ are the mean and standard deviation of the $s$ values on calibration set.
	    \State Define $b$ as the average of absolute error in calibration set.
	    \State Define the scaling factor $\sigma$ as a function of parameter $a$:
            \[ \sigma(a;\tilde{s}):= a*\tanh(\tilde{s})+b, \quad  0 \leq a \leq b\]
        \State Generate a list of candidate values for parameter a, denoted $A$, e.g. $A=\{0,b/100, 2*b/100, \ldots, b\}$.
		\For {each $a \in A$}
	    	\For {$repeat=1,2,\ldots,R$}
			    \State Randomly split the calibration into two halves, denoted as set $C_1, C_2$, which will be used as calibration set and test set in repeated cross-validation 
		        \State Use $C_1$ and scaling factor $\sigma(a)$ to construct conformal prediction intervals for $C_2$ under a specified normal level
		        \State Compute the average PI width $w(a;repeat)$
		        \State Divide set $C_2$ into four equal-sized subsets according to the prediction interval widths 
		        \State Compute the average of absolute coverage error on the four subsets in $C_2$, denoted as $\eta(a;repeat)$
			\EndFor
			\State Compute $\eta_m(a)$ and $\eta_s(a)$ as the average and standard deviation of $\eta(a;repeat)$ over $R$ repeats
			\State Compute $w_m(a)$ and $w_s(a)$ as the average and standard deviation of $w(a;repeat)$ over $R$ repeats
		\EndFor
		\State Let $S_1:=\{a \in A: \eta_m(a) \leq min(\eta_m(a))+\eta_s(\underset{a}{\operatorname{\argmin}}(\eta_m(a))/\sqrt{R} \}$
		\State Let $S_2:=\{a \in S_1: w_m(a) \leq min(w_m(a))+w_s(\underset{a}{\operatorname{\argmin}}(w_m(a))/\sqrt{R} \}$
		\State Compute $a_{opt} = median(\{a: a \in S_2 \})$
		\State Return the optimized scaling factor $\sigma(\tilde{s}):= a_{opt}*\tanh(\tilde{s})+b$
	\end{algorithmic} 
\end{algorithm}

\subsection{ML algorithm implementation and parameters}
\label{SI:Implementation}

\subsubsection{Random Forests}
The RF models were implemented using the scikit-learn python library \cite{pedregosa2011scikit} with a fixed set of hyperparameters: n\_estimators = 500, max\_features=0.33, min\_samples\_leaf=5, and default parameters otherwise. 

\subsubsection{Deep Neural Networks}
Both the DNN-dropout and DNN-multitask models are fully connected feed forward neural networks. They were implemented using Tensorflow \cite{abadi2016tensorflow} with the following structures for two groups of QSAR datasets:

\begin{table}[H]
\centering
\resizebox{0.8\columnwidth}{!}{
\begin{tabular}{l|l|l}
\hline
\textbf{ChEMBL and Simulation}                                                                  & \textbf{DNN-dropout}                                                                                                                                                                                                          & \textbf{DNN-multitask}                                                          \\ \hline \hline
\rowcolor[HTML]{DAE8FC} 
Number of nodes in output layers                                                                & 1                                                                                                                                                                                                                             & 20                                                                              \\ \hline
\rowcolor[HTML]{DAE8FC} 
Number of nodes in hidden layers                                                                & {[}1000, 1000, 100, 10{]}                                                                                                                                                                                                     & {[}1500, 1000, 500{]}                                                           \\ \hline
Hidden layer activation function                                                                & ReLU                                                                                                                                                                                                                          & ReLU                                                                            \\ \hline
\rowcolor[HTML]{DAE8FC} 
Dropout rates                                                                                   & {[}0.25, 0.25, 0.25, 0.25{]}                                                                                                                                                                                                  & {[}0.2, 0.2, 0.2{]}                                                             \\ \hline
\rowcolor[HTML]{DAE8FC} 
Random dropout during prediction                                                                & Yes                                                                                                                                                                                                                           & No                                                                              \\ \hline
\rowcolor[HTML]{DAE8FC} 
\begin{tabular}[c]{@{}l@{}}Number of forward passes \\ for predicting one molecule\end{tabular} & 100                                                                                                                                                                                                                           & 1                                                                               \\ \hline
\rowcolor[HTML]{DAE8FC} 
Optimizer                                                                                       & \begin{tabular}[c]{@{}l@{}}SGD, \\ momentum=0.9, \\ nesterov=True\end{tabular}                                                                                                                                                & \begin{tabular}[c]{@{}l@{}}SGD, \\ momentum=0.9, \\ nesterov=False\end{tabular} \\ \hline
\rowcolor[HTML]{DAE8FC} 
Batch size                                                                                      & $n_{training}$*0.15                                                                                                                                                                                                              & $n_{training}$/20                                                                  \\ \hline
\rowcolor[HTML]{DAE8FC} 
Learning rate                                                                                   & \begin{tabular}[c]{@{}l@{}}Cyclical annealing: starts with 0.005, \\ decreases by 40\% every 200 epochs, \\ set back to initial value 0.005 after \\ each 1000 epochs, and repeat the \\ annealing process again\end{tabular} & 0.001                                                                           \\ \hline
\rowcolor[HTML]{DAE8FC} 
Epochs                                                                                          & 4000                                                                                                                                                                                                                          & 500                                                                             \\ \hline
\rowcolor[HTML]{DAE8FC} 
Early stopping                                                                                  & patience=300                                                                                                                                                                                                                  & patience=30                                                                     \\ \hline
L2 regularization penalty                                                                       & 0.00005                                                                                                                                                                                                                       & 0.00005                                                                         \\ \hline
\end{tabular}
}
\end{table}

\begin{table}[H]
\centering
\resizebox{0.8\columnwidth}{!}{
\begin{tabular}{l|l|l}
\hline
\textbf{Kaggle}                                                                                 & \textbf{DNN-dropout}                                                         & \textbf{DNN-multitask}                                                       \\ \hline \hline
\rowcolor[HTML]{DAE8FC} 
Number of nodes in output layers                                                                & 1                                                                            & 50                                                                           \\ \hline
Number of nodes in hidden layers                                                                & {[}4000, 2000, 1000, 1000{]}                                                 & {[}4000, 2000, 1000, 1000{]}                                                 \\ \hline
Hidden layer activation function                                                                & ReLU                                                                         & ReLU                                                                         \\ \hline
Dropout rates                                                                                   & {[}0.25, 0.25, 0.25, 0.1{]}                                                  & {[}0.25, 0.25, 0.25, 0.1{]}                                                  \\ \hline
\rowcolor[HTML]{DAE8FC} 
Random dropout during prediction                                                                & Yes                                                                          & No                                                                           \\ \hline
\rowcolor[HTML]{DAE8FC} 
\begin{tabular}[c]{@{}l@{}}Number of forward passes \\ for predicting one molecule\end{tabular} & 100                                                                          & 1                                                                            \\ \hline
Optimizer                                                                                       & \begin{tabular}[c]{@{}l@{}}SGD, \\  momentum=0.9, \\nesterov=False\end{tabular} & \begin{tabular}[c]{@{}l@{}}SGD, \\ momentum=0.9, \\nesterov=False\end{tabular} \\ \hline
Batch size                                                                                      & min(128,$n_{training}$/20)                                                      & min(128,$n_{training}$/20)                                                      \\ \hline
Learning rate                                                                                   & 0.001                                                                        & 0.001                                                                        \\ \hline
Epochs                                                                                          & 500                                                                          & 500                                                                          \\ \hline
Early stopping                                                                                  & patience=50                                                                  & patience=50                                                                  \\ \hline
L2 regularization penalty                                                                       & 0.00005                                                                      & 0.00005                                                                      \\ \hline
\end{tabular}
}
\end{table}

\subsubsection{Gradient Boosting}
The python library LightGBM \cite{ke2017lightgbm} is used to implement the gradient boosting models. We use one set of LightGBM-related hyperparameters for all datasets:

\begin{table}[H]
\centering
\begin{tabular}{l|l}
\hline
num\_leaves       & 64         \\ \hline
objective         & regression \\ \hline
metric            & mse        \\ \hline
bagging\_freq     & 1          \\ \hline
bagging\_fraction & 0.7        \\ \hline
feature\_fraction & 0.7        \\ \hline
learning\_rate    & 0.01       \\ \hline
num\_iterations   & 1500       \\ \hline
boosting\_type    & gbdt       \\ \hline
\end{tabular}
\end{table}

\subsection{Hyperparameter tuning for prediction uncertainty score methods}
\label{SI:tuning}

As mentioned in Section \ref{sec: Methods}, we designed novel algorithms, DNN-multitask and GB-tail, to compute the prediction uncertainty scores for Deep Neural Networks and Gradient Boosting respectively. 

The hyperparameters in DNN-multitask method are number of output nodes (K) and random omissions probability (p) for creating sparse multitask training labels. We evaluate 12 combinations of $(K, p)$ pairs: 
\[ K \in \{10, 20, 50, 100\}, \quad p \in \{0, \exp^{-1}, 0.6\}  \]
And for GB-tail method, the only hyperparameter $w$ is the proportion of ``tail iterations'' in a trained Gradient Boosting model used for computing the prediction uncertainty. We tried a list of values:
\[ w \in \{0.2, 0.4, 0.6, 0.8\} \]

\begin{figure}[H]
\centering
\caption{Hyperparameter tuning of prediction uncertainty score methods for Deep Neural Networks and Gradient Boosting}
\includegraphics[width=0.9\textwidth]{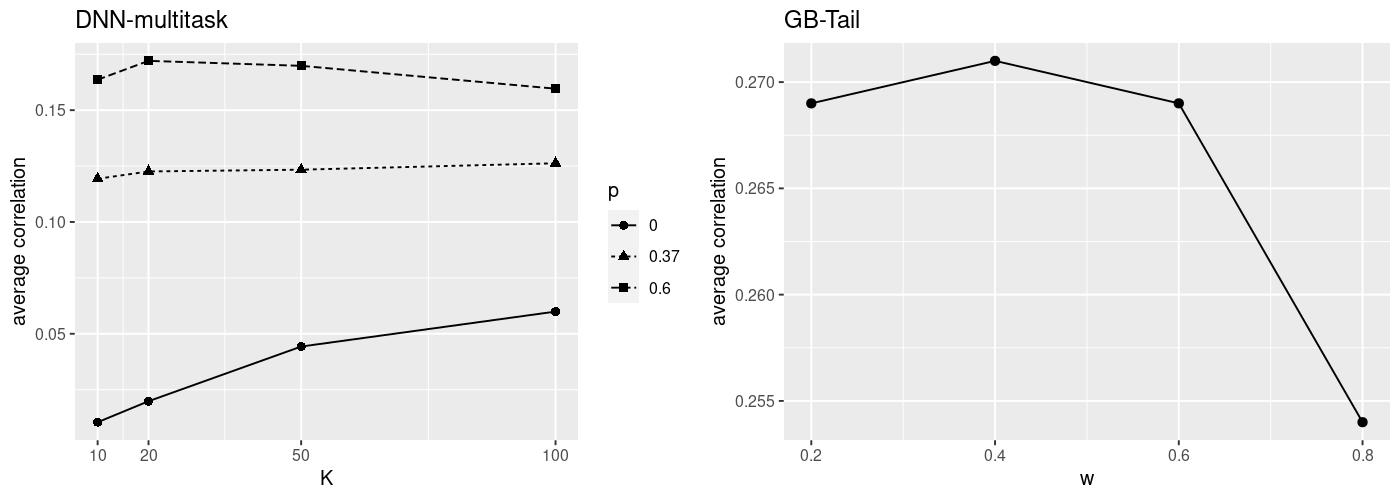}
\label{fig:hyperparameterTuning}
\end{figure}

We use the datasets in simulation case 2 to evaluate the prediction uncertainty scores from DNN-multitask and GB-tail methods with different hyperparameter settings. The average Spearman's rank correlation between prediction uncertainty score and absolute error in validation set was calculated across all datasets and repeats.  
Although the prediction uncertainty score is not directly predicting the absolute error, we hope that larger absolute errors only happen with higher prediction uncertainty score, so larger correlation is preferred. 
Figure \ref{fig:hyperparameterTuning} shows the comparison of average correlation under various hyperparameters for both methods. 
For DNN-multitask method, larger omission percentage (p=0.6), which means more sparsity in multitask training label, is preferred; and moderate number of output nodes (K = 20 or 50) perform slightly better than the more extreme choices. 
For GB-tail method, the correlation is not sensitive with small and moderate values of $w$, but decreases sharply when using larger $w$. We concludes that it is safer to choose smaller value $w = 0.2$, so that GB-tail method only use the iterations after most data gets stable predictions over the iteration sequence.

\subsection{Simulation Settings}
\label{SI:SimulationDetails}

For each ChEMBL dataset, we first build an RF model, and use the out-of-bag predictions $\hat{y} = f(X)$ as the true activity of a molecule with descriptor $X$. Let $sd_{oob}$ being the standard deviation of the residuals. 

Simulate activity $\tilde{y}$ with homoscedastic or heteroscedastic noise $e \sim \mathbf{N}(0, \sigma^2)$: $\tilde{y} = f(X) + e $. 

\begin{itemize}
    \item Case 1: Homoscedastic variance: $\sigma = 0.8*sd_{oob}$
    \item Case 2: Heteroscedastic variance: $\sigma = g(X)$
            The standard deviation of error term, $g(X)$, is a S-shape monotone increasing function of $f(X)$. 
            First scale $f(X_i)$ between -1 and 1 within each dataset, and denote the scaled f as u, i.e.
            \[u(X):= 2*(f(X)-f_{min})/(f_{max}-f_{min})-1,\]
            where $f_{min}$ and $f_{max}$ is the minimum and maximum of $f(X)$ in a data set respectively. 
            Then define 
            \[ g(X) = 0.1 + exp(3*u(X)/\big(exp(3*u(X))+1\big)  \]
            Finally scale the $g(X)$ within each dataset, so that the average of $g(X)$ is the same with the constant standard deviation in homoscedastic case for the same dataset.
\end{itemize}

\subsection{Additional Figures}
\label{SI:AdditionalFigures}

\begin{figure}[H]
\centering
\caption{The scales of prediction uncertainty scores from different methods. (Simulation 1)}
\includegraphics[width=\textwidth]{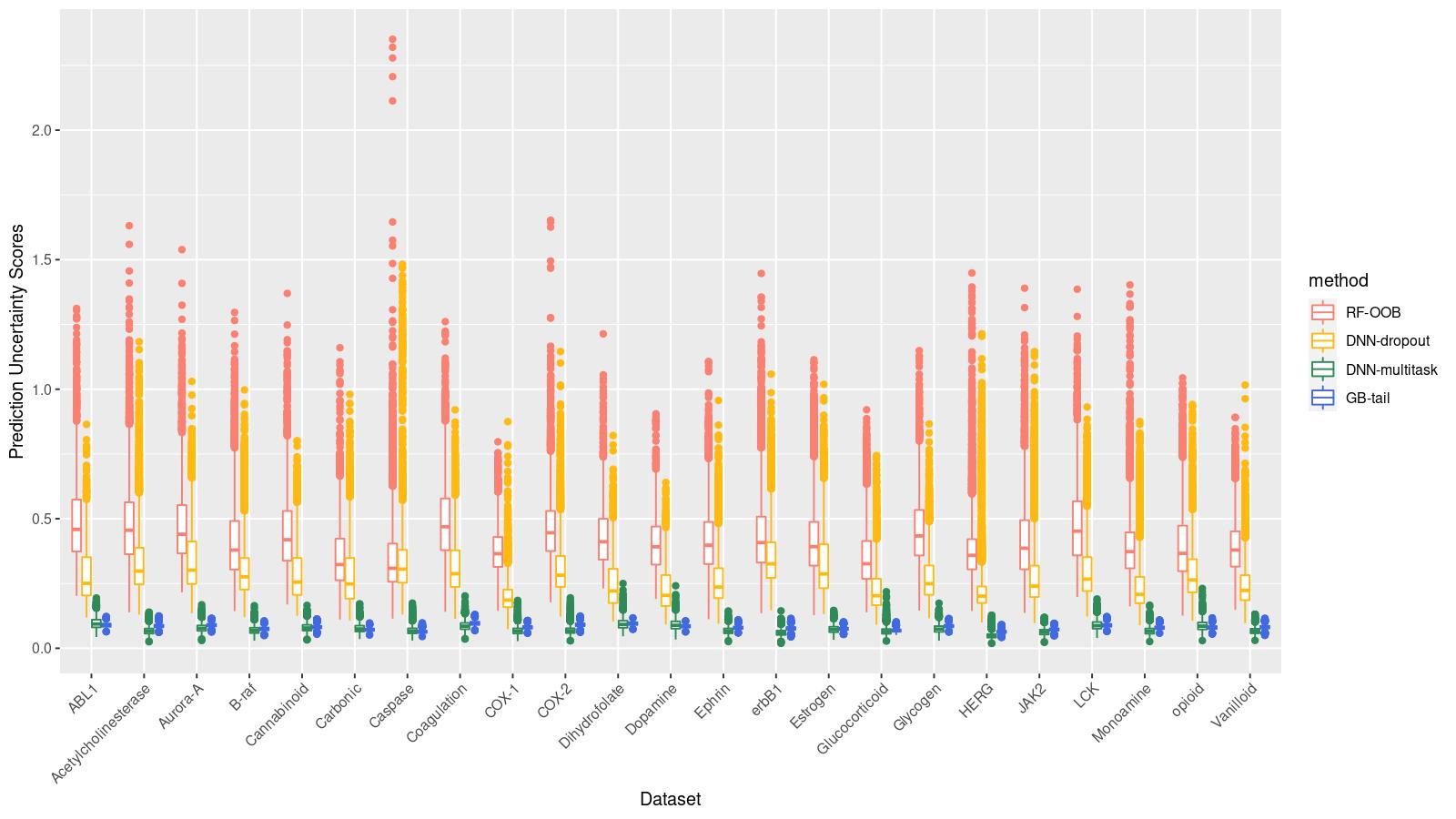}
\label{fig:sim1_PredSD}
\end{figure}

\begin{figure}[H]
\centering
\caption{Prediction interval widths comparison for nominal level 90\%  (Simulation 1)}
\includegraphics[width=\textwidth]{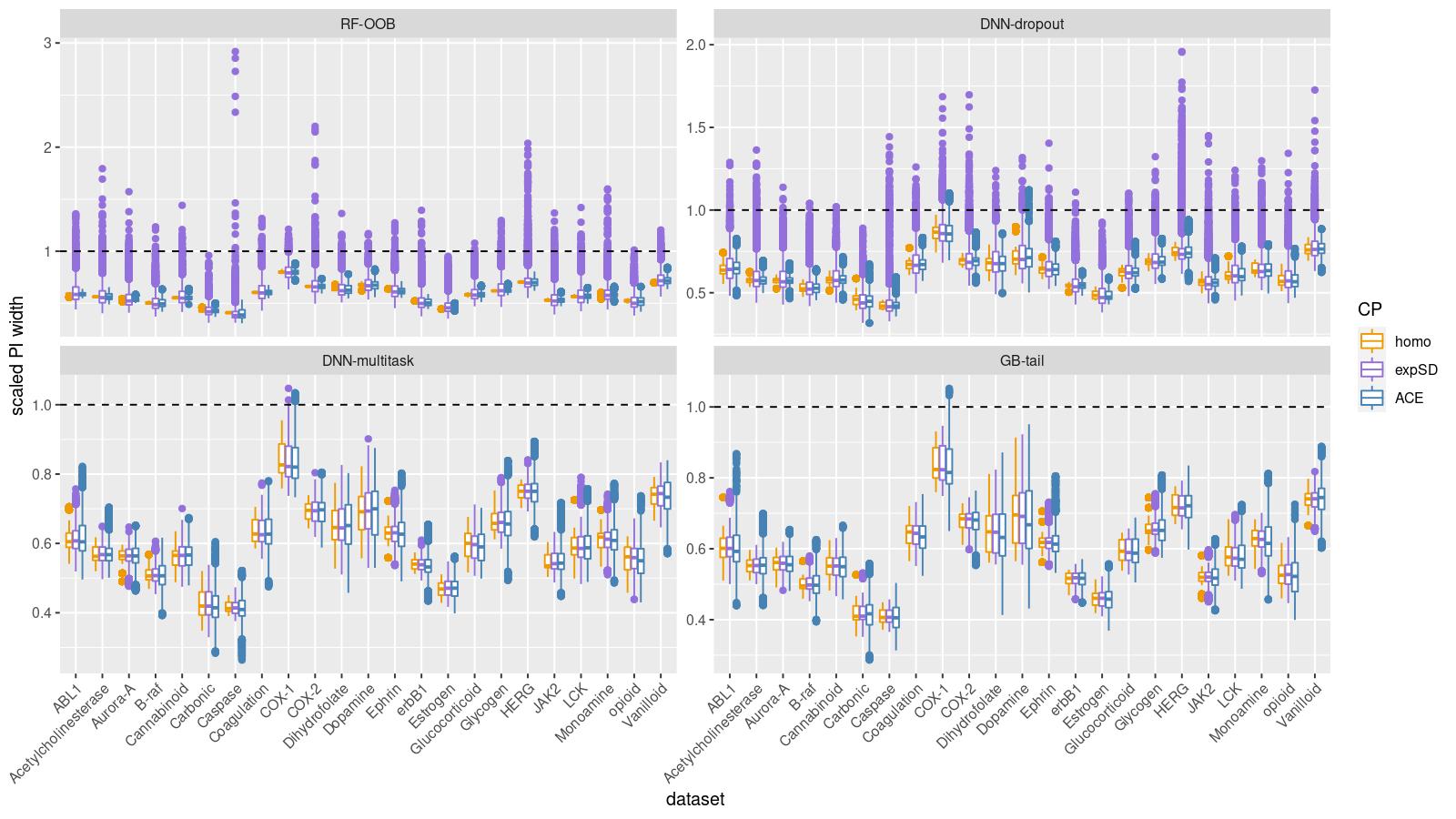}
\label{fig:sim1_PIwidth_dataset}
\end{figure}

\begin{figure}[H]
\centering
\caption{Test set coverage of subgroups defined by prediction interval width, at nominal level 70\%}
\includegraphics[width=\textwidth]{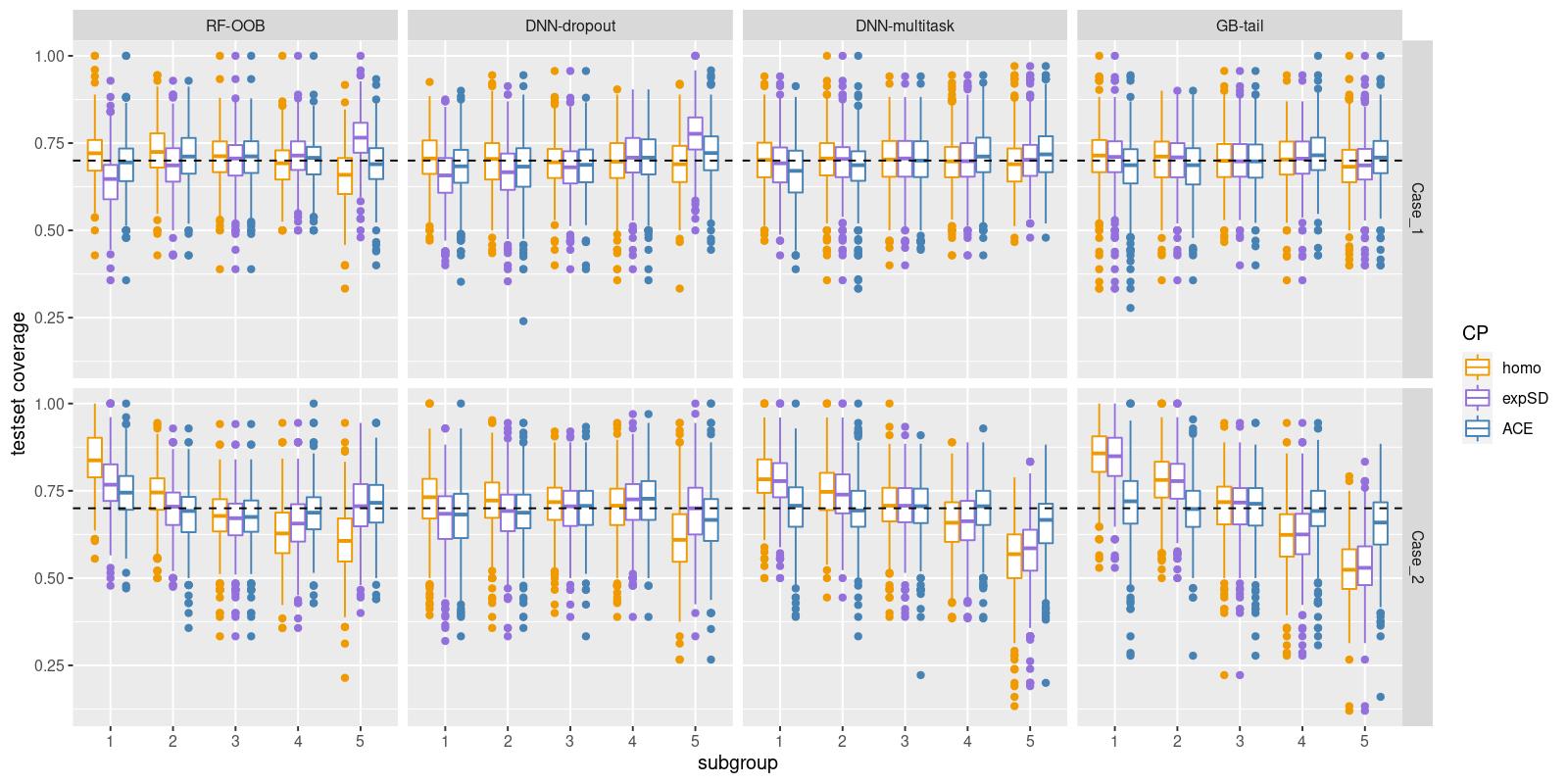}
\label{fig:sim_subGroup_70}
\end{figure}

\begin{figure}[H]
\centering
\caption{Average prediction interval widths of subgroups defined by prediction interval widths, at nominal level 70\%. The prediction intervals are scaled by ``no-model'' interval width.}
\includegraphics[width=\textwidth]{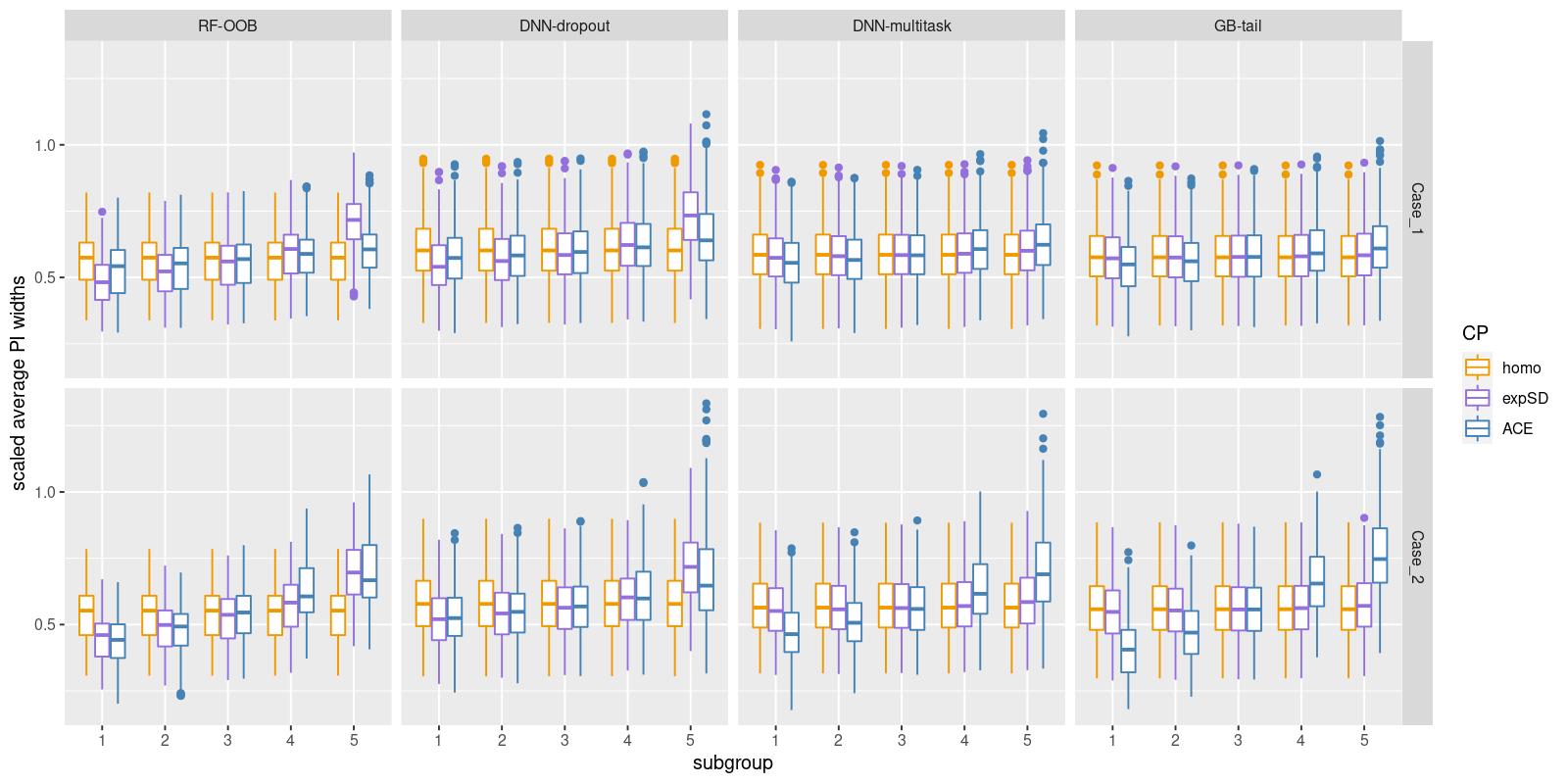}
\label{fig:sim_subGroup_70_avgPIwidth}
\end{figure}

\begin{figure}[H]
\centering
\caption{Test set coverage of subgroups defined by prediction interval width, at nominal level 90\%}
\includegraphics[width=\textwidth]{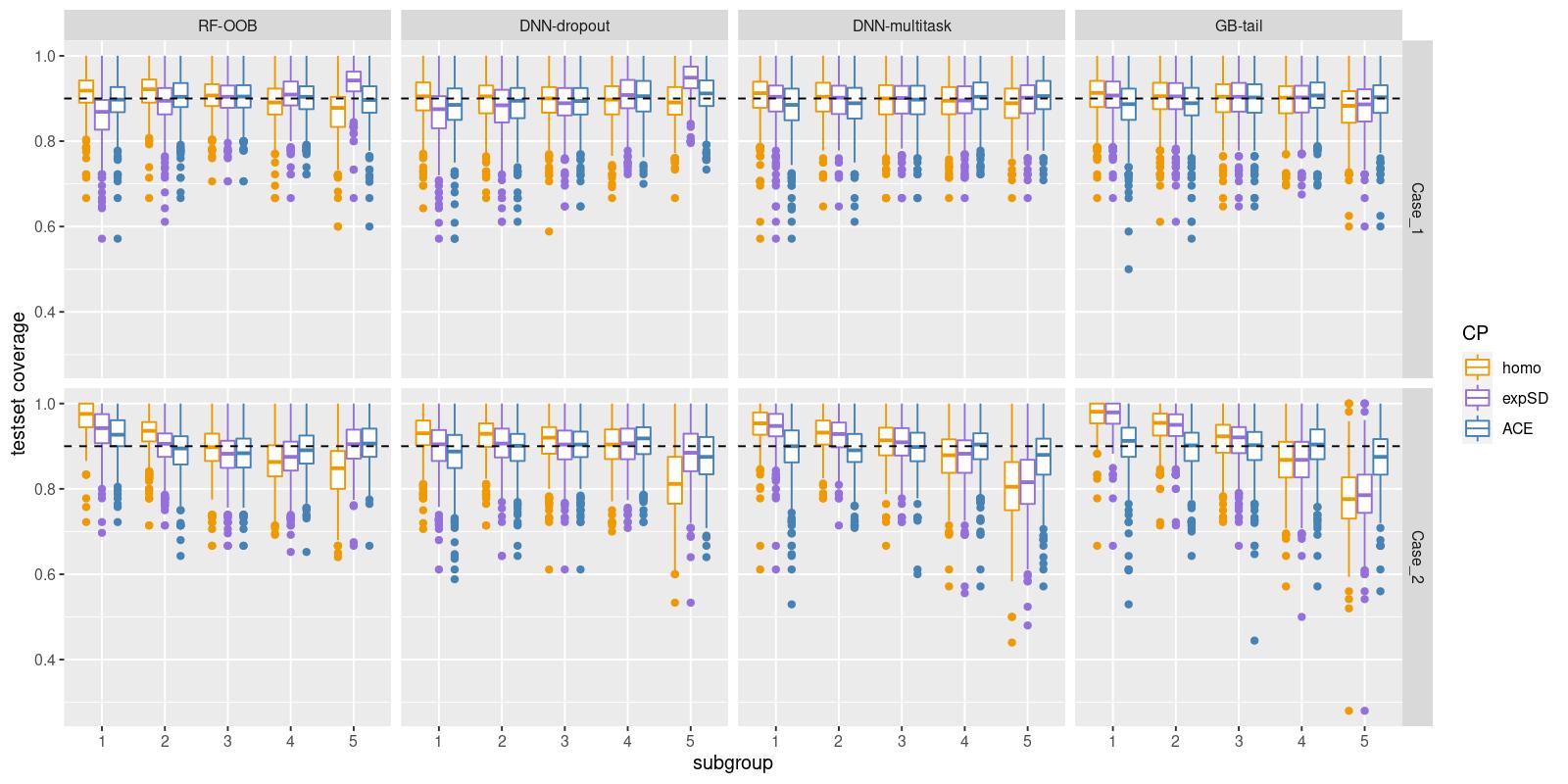}
\label{fig:sim_subGroup_90}
\end{figure}

\begin{figure}[H]
\centering
\caption{Average prediction interval widths of subgroups defined by prediction interval widths, at nominal level 90\%. The prediction intervals are scaled by ``no-model'' interval width.}
\includegraphics[width=\textwidth]{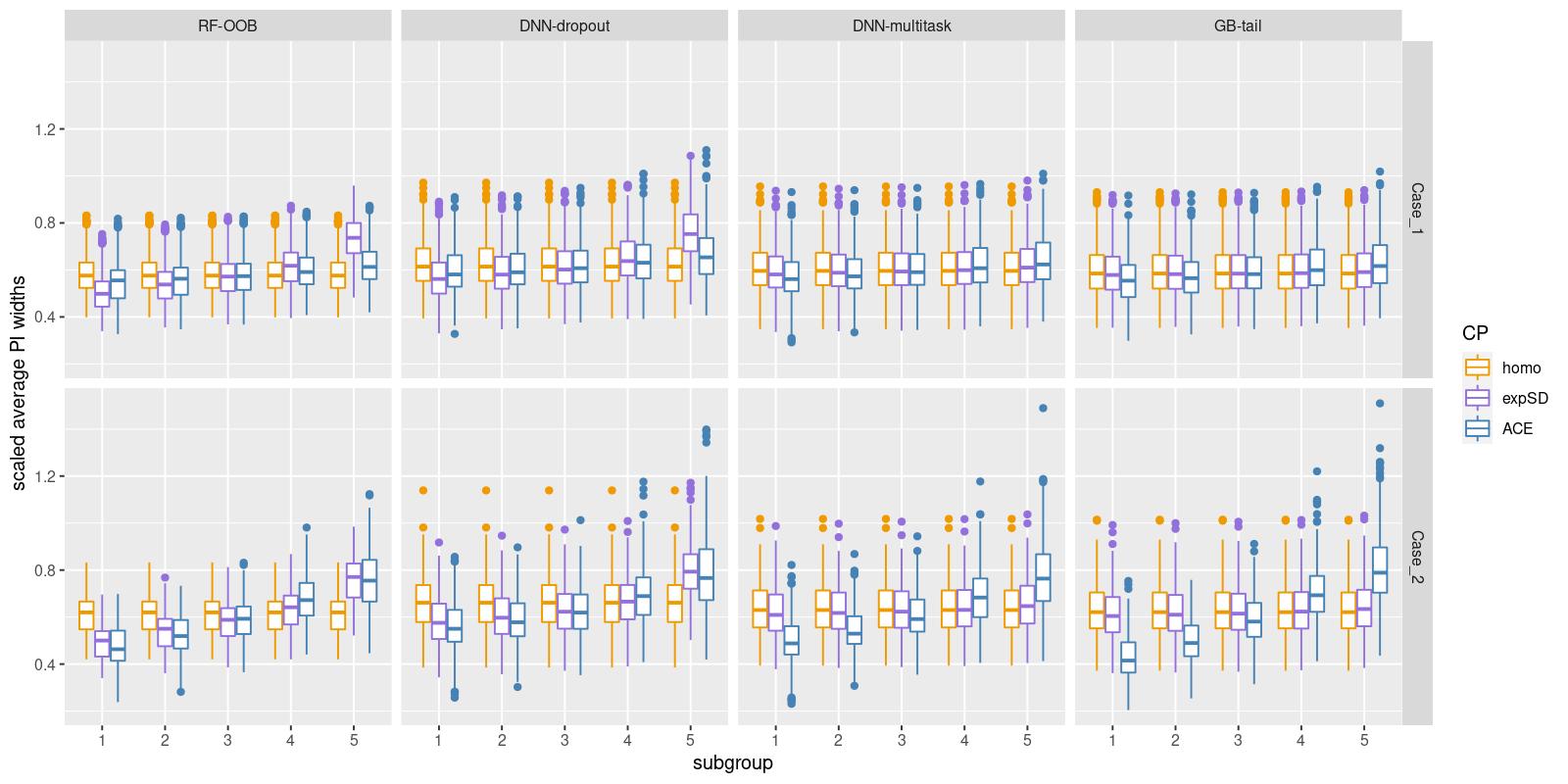}
\label{fig:sim_subGroup_90_avgPIwidth}
\end{figure}

\nolinenumbers

\newpage
\bibliographystyle{achemso99}
\bibliography{ref}

\end{document}